\documentclass{aa}  

\usepackage{graphicx}
\bibpunct{(}{)}{;}{a}{}{,}
\usepackage{txfonts}
\usepackage{amssymb}
\usepackage{multirow}
\usepackage{float}
\begin{document} 

   \title{Shadows and cavities in protoplanetary disks: \\ HD163296, HD141569A, and HD150193A in polarized light\thanks{Based on observations collected at the European Organisation for Astronomical Research in the Southern Hemisphere, Chile, under program number 089.C-0611(A).}\ \thanks{This paper makes use of the following ALMA data: ADS/JAO.ALMA\#ADS/JAO.ALMA\#2011.0.00010.SV. ALMA is a partnership of ESO (representing its member states), NSF (USA) and NINS (Japan), together with NRC (Canada) and NSC and ASIAA (Taiwan), in cooperation with the
Republic of Chile. The Joint ALMA Observatory is operated by ESO, AUI/NRAO and NAOJ.} }

  \author{A. Garufi
          \inst{1}
          \and
          S.P. Quanz
          \inst{1}
          \and
          H.M. Schmid
          \inst{1}
          \and
          H. Avenhaus
          \inst{1}
          \and
          E. Buenzli
          \inst{2}
          \and
          S. Wolf
          \inst{3}
          }

   \institute{Institute for Astronomy, ETH Zurich, Wolfgang-Pauli-Strasse 27, CH-8093 Zurich, Switzerland\\
              \email{antonio.garufi@phys.ethz.ch}
         \and
            Max Planck Institute for Astronomy, K\"{o}nigstuhl 17, D-69117, Heidelberg, Germany
         \and
         University of Kiel, Institute of Theoretical Physics and Astrophysics, Leibnizstrasse 15, 24098 Kiel, Germany  
             }

  \date{Received ...; accepted ...}
  
 
  \abstract
   {The morphological evolution of dusty disks around young (a
few Myr old) stars is pivotal for a better understanding of planet formation. Since both dust grains and the global disk geometry evolve on short timescales, high-resolution imaging of a sample of objects may provide important indications about this evolution.}
   {We enlarge the sample of protoplanetary disks imaged in polarized light with high-resolution imaging ($\lesssim 0.2\arcsec$) by observing the Herbig Ae/Be stars HD163296, HD141569A, and HD150193A. We combine our data with previous datasets to understand the larger context of their morphology.}
   {Polarimetric differential imaging is an attractive technique with which to image at near-IR wavelengths a significant fraction of the light scattered by the circumstellar material. The unpolarized stellar light is canceled out by combining two simultaneous orthogonal polarization states. This allowed us to achieve an inner working angle and an angular resolution as low as  $\sim0.1\arcsec$.}
   {We report a weak detection of the disk around HD163296 in the $H$ and $K_{\rm S}$ band. The disk is resolved as a broken ring structure with a significant surface brightness drop inward of $0.6\arcsec$. No sign of extended polarized emission is detected from the disk around HD141569A and HD150193A.}
   {We propose that the absence of scattered light in the inner $0.6\arcsec$ around HD163296 and the non-detection of the disk around HD150193A may be due to similar geometric factors. Since these disks are known to be flat or only moderately flared, self-shadowing by the disk inner wall is the favored explanation. We show that the polarized brightness of a number of disks is indeed related to their flaring angle. Other scenarios (such as dust grain growth or interaction with icy molecules) are also discussed. On the other hand, the non-detection of HD141569A is consistent with previous datasets that revealed a huge cavity in the dusty disk.}


\authorrunning{Garufi et al.}

\titlerunning{Shadows and cavities in protoplanetary disks}

   \maketitle
%

\section{Introduction}
To better comprehend the mechanisms that govern planet formation, an accurate knowledge of the morphology of circumstellar disks is needed. Since a high degree of radial/azimuthal asymmetries is to be expected for these disks, a comprehensive dataset must include, among other data, high-resolution imaging. 

Moderate- to high-resolution maps of the dusty disks around young stars have been obtained at submm \citep[e.g.,][]{Andrews2009} and near-IR wavelengths. The latter, in particular, can be either performed with total scattered light \citep[e.g.,][]{Grady2009} or polarized light \citep[e.g.,][]{Muto2012}, which are often complementary rather than analogous. In fact, scattered-light images are usually obtained by subtracting a point-spread function (PSF) reference star and with the aid of a coronagraph and can, thus, trace the dust in the outer ($\gtrsim 1\arcsec$) part of the disk surface. On the other hand, polarized-light observations can achieve an inner working angle (IWA) as small as $0.1\arcsec$, but are often sensitivity limited to an outer working angle (OWA) of $1\arcsec-1.5\arcsec$.

Recently, polarimetric differential imaging \citep[PDI, e.g.,][]{Apai2004, Quanz2011, Hashimoto2012} observations of circumstellar disks were shown to be a fundamental tool for resolving peculiar structures in disks at radii of a few tens of AU where planet formation is thought to occur. This technique relies on the fact that direct stellar light, unlike a significant fraction of the scattered light from the disk, is unpolarized. PDI effectively suppresses residual speckles from the atmospheric seeing corrections of an AO system and diffraction effects from the optics. 

\begin{table*}
      \caption[]{Summary of observations. The columns list the
object name, filter name, detector integration time (DIT), number of integrations per retarder plate position (NDIT) multiplied by integrations per dither position (NINT), number of dither positions, total integration time (TIT) per filter, and average airmass, optical seeing, and coherence time. Note that TIT per filter is obtained from the sum of all (DIT $\times$ NDIT $\times$ NINT $\times$ dither pos.) multiplied by the four retarder plate positions.}
         \label{Settings}
     $$ 
         \begin{tabular}{ccccccccc}
            \hline
            \hline
            \noalign{\smallskip}
            Star & Filter & DIT (s) & NDIT $\times$ NINT & Dither pos. & TIT (s) & <Airmass> & <Seeing> & <$\tau_0$> (ms) \\
            \hline
           \noalign{\smallskip}
            \multirow{4}{*}{HD163296} & $NB \ 1.64$ & 0.3454 & 15 $\times$ 3 & 3 & 186.5 & 1.06 & 1.05$\arcsec$ & 26 \\
            & $H$ & 0.3454 & 140 $\times$ 6 & 3 & 3481.6 & 1.03 & 1.02$\arcsec$ & 27 \\
           & $NB \ 2.17$ & 0.3454 & 15 $\times$ 1 & 3 & 62.2 & 1.22 & 0.81$\arcsec$ & 33 \\
           & $K_{\rm s}$ & 0.3454 & 100 $\times$ 3 & 3 & 1243.4 & 1.16 & 0.97$\arcsec$ & 27 \\
            \hline
            \noalign{\smallskip}
            \multirow{5}{*}{HD141549A} & \multirow{3}{*}{$NB \ 1.64$} & 0.7 & 15 $\times$ 1 & 3 & \multirow{3}{*}{486} & \multirow{3}{*}{1.11} & \multirow{3}{*}{1.40$\arcsec$} & \multirow{3}{*}{17} \\
            & & 1 & 10 $\times$ 1 & 3 & & & & \\
            & & 2 & 10 $\times$ 1 & 3 & & & & \\
            \noalign{\smallskip}
            & \multirow{2}{*}{$H$} & 0.5 & 90 $\times$ 5 & 3 & \multirow{2}{*}{5400} & \multirow{2}{*}{1.08} & \multirow{2}{*}{1.66$\arcsec$} & \multirow{2}{*}{14} \\
            & & 3 & 75 $\times$ 1 & 3 & & & & \\
           \noalign{\smallskip}
            \hline
            \noalign{\smallskip}
            \multirow{4}{*}{HD150193A} & $NB \ 1.64$ & 0.3454 & 20 $\times$ 3 & 3 & 248.7 & 1.26 & 0.62$\arcsec$ & 44 \\
            & $H$ & 0.3454 & 140 $\times$ 8 & 3 & 4642.2 & 1.21 & 0.58$\arcsec$ & 45 \\
           & $NB \ 2.17$ & 0.3454 & 20 $\times$ 2 & 3 & 165.8 & 1.10 & 1.43$\arcsec$ & 17 \\
           & $K_{\rm s}$ & 0.5 & 95 $\times$ 2 & 6 & 2280 & 1.09 & 1.27$\arcsec$ & 18 \\
           \noalign{\smallskip}
                                   \hline
            \hline
            \noalign{\smallskip}
         \end{tabular}
     $$ 

   \end{table*}

Scattered-light images of inclined circumstellar disks are expected to show azimuthal asymmetries due to anisotropic scattering by the dust grains. These anisotropies depend on the grain properties. The scattering function for larger grains is indeed more forward-peaking \citep[e.g.,][]{Mulders2013}. The amount of this is difficult to retrieve, however. On top of this effect, scatterers polarize the light by a fraction that depends on the nature of the scattering particles (e.g., composition, compactness) and on the scattering angle. In PDI images, these two effects are not easy to distinguish if complementary scattered-light images are unavailable.

In addition to the dust grain properties, the global disk geometry can also considerably influence scattered-light observations. Flared geometries have been invoked to explain the high far-IR excess in the spectral energy distribution (SED) of young stars. Herbig Ae/Be disks are typically classified into flared (group I) and flat (group II) objects, based upon the shape of their IR excess \citep{Meeus2001}. The disk scale height can also affect the amount of light that is scattered by the disk surface. 

The geometry of the inner few AU of disks can also play an important role for scattered-light images. In fact, the inner wall can intercept a significant fraction of the stellar radiation and cast a shadow on the outer disk \citep[e.g.,][]{Dullemond2001}. The scale height of these walls is known to be variable, as suggested by the typically observed erratic near-IR magnitudes. An anticorrelation between the near-IR and the far-IR excess has been found in some objects \citep[e.g.,][]{Espaillat2011}, which supports this scenario.


In this paper we present new PDI observations of HD163296, HD141569A, and HD150193A obtained with the AO-assisted high-resolution near-IR NAOS/CONICA instrument \citep[NACO][]{Lenzen2003, Rousset2003} at the Very Large Telescope (VLT). These three sources are all known to host circumstellar disks from both scattered-light and thermal images.

HD163296 is an A1 star at $d=122$ pc \citep{vandenAncker1998} that appears to be physically unassociated with any known star-forming region \citep{Finkenzeller1984}. This source hosts a large circumstellar disk ($R \simeq 500$ AU) that has been inferred to be quite evolved. The gaseous disk is known to be twice as large as the dusty disk as traced by (sub-)mm imaging \citep{Isella2007, deGregorio-Monsalvo2013}. Scattered-light images \citep{Grady2000} trace small dust grains roughly out to the outer edge of the gaseous disk. {Taken together, these observations have} been interpreted as a sign of grain growth and inward migration for larger dust grains. However, no sign of dust or gas inner clearing has been observed at radii $r>25$ AU \citep{deGregorio-Monsalvo2013}. Interferometric near-IR observations show that a significant fraction of emission is originating at the theoretical dust sublimation radius \citep{Benisty2010}. This object is also one of the few for which the location of the CO snowline has been inferred \citep[155 AU,][]{Qi2011}. Finally, multi-epoch coronagraphic images of the source reveal variable scattered light from the outer ($>2.9\arcsec$) disk \citep{Wisniewski2008}, whereas the stellar V magnitude is rather stable \citep{Herbst1999}. A possible explanation for this is a variable scale height of the disk inner wall that is also responsible for near-IR variability \citep{Sitko2008}. 

The A0 star HD141569A \citep[$d=99$ pc,][]{vandenAncker1998} is part of a physical triple system with two M-dwarf companions at $9\arcsec$ from the primary \citep{Weinberger2000}. Its large ($R \simeq 400$ AU) asymmetric circumstellar disk has been imaged in scattered light by \citet{Augereau1999}, \citet{Weinberger1999}, \citet{Mouillet2001}, and \citet{Clampin2003}, who resolved a prominent spiral structure that was ascribed to the gravitational interaction with the two companions. Mid-IR imaging by \citet{Marsh2002} revealed a depression in the optical depth in the inner 30 AU, consistent with the near-IR deficit in the SED \citep{Sylvester1996}. 

HD150193A is an A2 star at $d=150$ pc \citep{vandenAncker1997} that is physically associated with a K4 star \citep{Bouvier2001} at $1.1\arcsec$ (HD150193B). A spatially unresolved disk ($R<250$ AU) around the primary star was detected by \citet{Mannings1997} at 2.6 mm. The disk was also imaged in scattered light at near-IR wavelengths by \citet{Fukagawa2003}. They ascribed a clear asymmetry in the disk structure to the interaction with HD150193B, where no disk was detected. Near-IR polarimetric images \citep{Hales2006} revealed an unresolved structure with high polarization and suggested that a large amount of polarizing material is in the line of sight toward the star \citep[in agreement with][]{Whittet1992}. 
 
The paper is organized as follows: first we describe in Sect.\,\ref{Observations} the observing conditions and data reduction. Then we present
in Sect.\,\ref{PDI_results} the results of our observations. Finally, in Sects.\,\ref{Discussion_HD163296} and\,\ref{Discussion_nondetection} we discuss the favored morphologies for all three sources. We conclude in Sect.\,\ref{Conclusions}.


\section{Observations and data reduction}\label{Observations}
HD163296, HD141569A, and HD150193A were observed between 2012 July 23 and 25 with VLT/NACO in PDI mode, consisting of AO-assisted high-resolution imaging polarimetry with a Wollaston prism and a rotatable half-wave plate for beam exchange. These sources were observed in the context of a small survey of six young stars, where the other objects (HD169142, HD142527, and SAO206462) were discussed by \citet{Quanz2013}, \citet{Avenhaus2014}, and \citet{Garufi2013}. 

We used the NACO SL27 camera (pixel scale = 27 mas pixel$^{-1}$) in \textit{HighDynamic} mode and readout in \textit{Double RdRstRd} mode. The targets were observed in $H$ and $K_{\rm S}$ bands, as well as in the respective narrow-band filters ($NB\ 1.64$ and $NB\ 2.17$). Since no photometric standard star was observed, narrow-band filters were necessary to obtain unsaturated exposures for photometric calibration. Observing conditions were photometric with average seeing varying from 0.58$\arcsec$ to 1.66$\arcsec$ (see Table \ref{Settings} for observing settings and conditions). 

The data were reduced with the procedure described in \citet{Avenhaus2014}. In addition to dark current, flat field, and bad pixel correction, this method consists of ($i$) performing a row-by-row subtraction to compensate for a nonstatic readout noise, ($ii$) determining the center of the stellar profile by fitting a two-dimensional Gaussian to the PSF, ($iii$) extracting ordinary and extraordinary beams from each image, upscaling the images by a factor 3 and aligning them on top of each other. The images thus obtained have a $3.24\arcsec \times 3.24\arcsec$ field of view. Instrumental polarization and cross-talk effects \citep{Witzel2010, Quanz2011} were considered by implementing the correction outlined in the Appendix of \citet{Avenhaus2014}. However, this approach requires quantifying all instrumental effects directly from the data. Since the targets of this paper show only a marginal or no polarized signal, some parameters (e.g., the relative efficiency between the Stokes parameters) were fixed to the values found for other objects \citep[see Table C1 of][]{Avenhaus2014}.

The photometric calibration (only accurate to $\sim40\%$) was performed as described by \citet{Quanz2011}. To do that, we assumed that the sources have the same magnitude in $H$ and $K_{\rm S}$ as in $NB\ 1.64$ and $NB\ 2.17$ . We converted the pixel-by-pixel count rates of the narrow-band images into surface brightness using the 2MASS magnitudes \citep{Cutri2003} and applied a scaling factor to the $H$ and $K_{\rm S}$ images taking into account the different transmission curves and exposure times. However, other uncertainties may derive from the variable near-IR flux of these sources \citep{Sitko2008, Pogodin2012}.

\begin{figure}
   \centering
 \includegraphics[width=4.4cm]{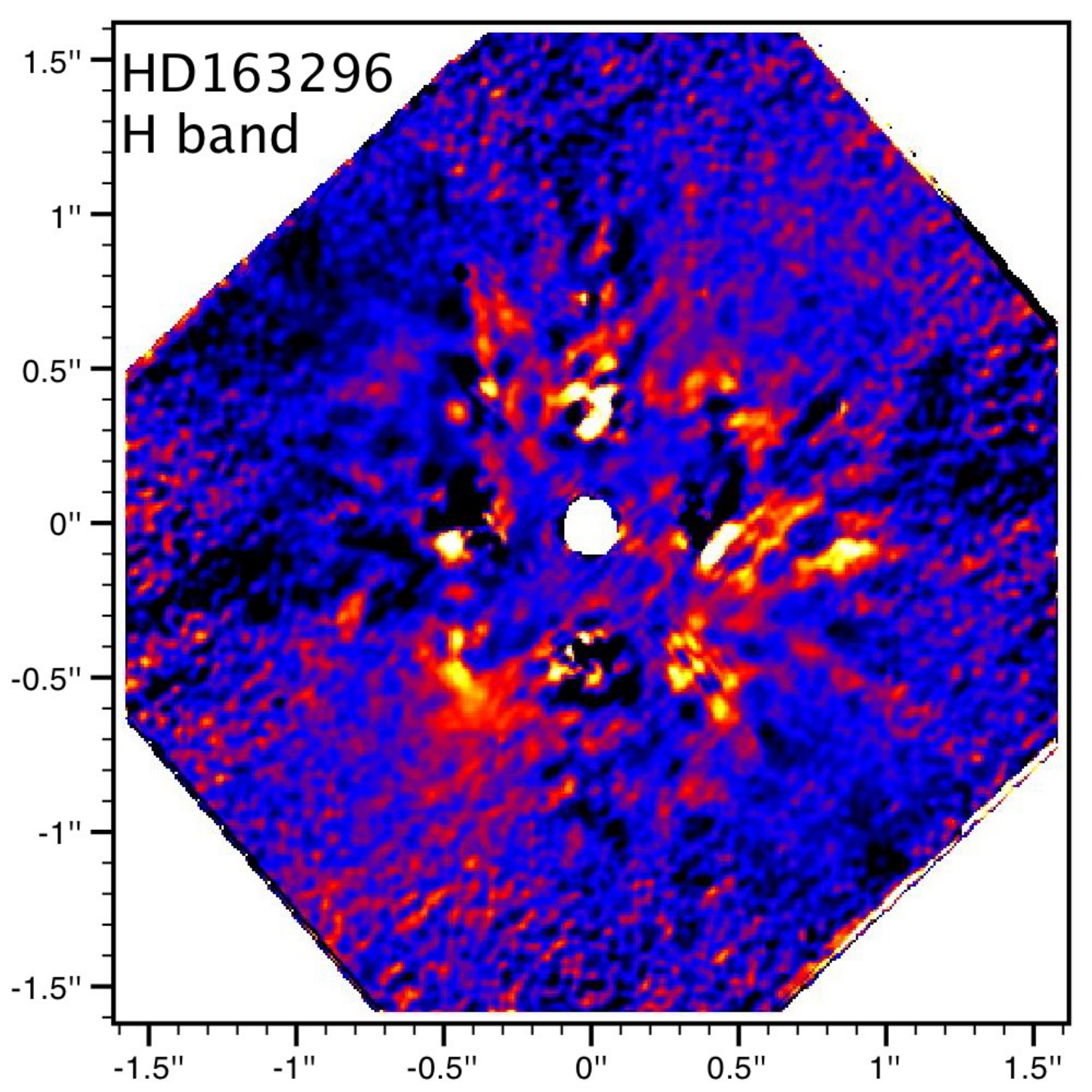}
\includegraphics[width=4.4cm]{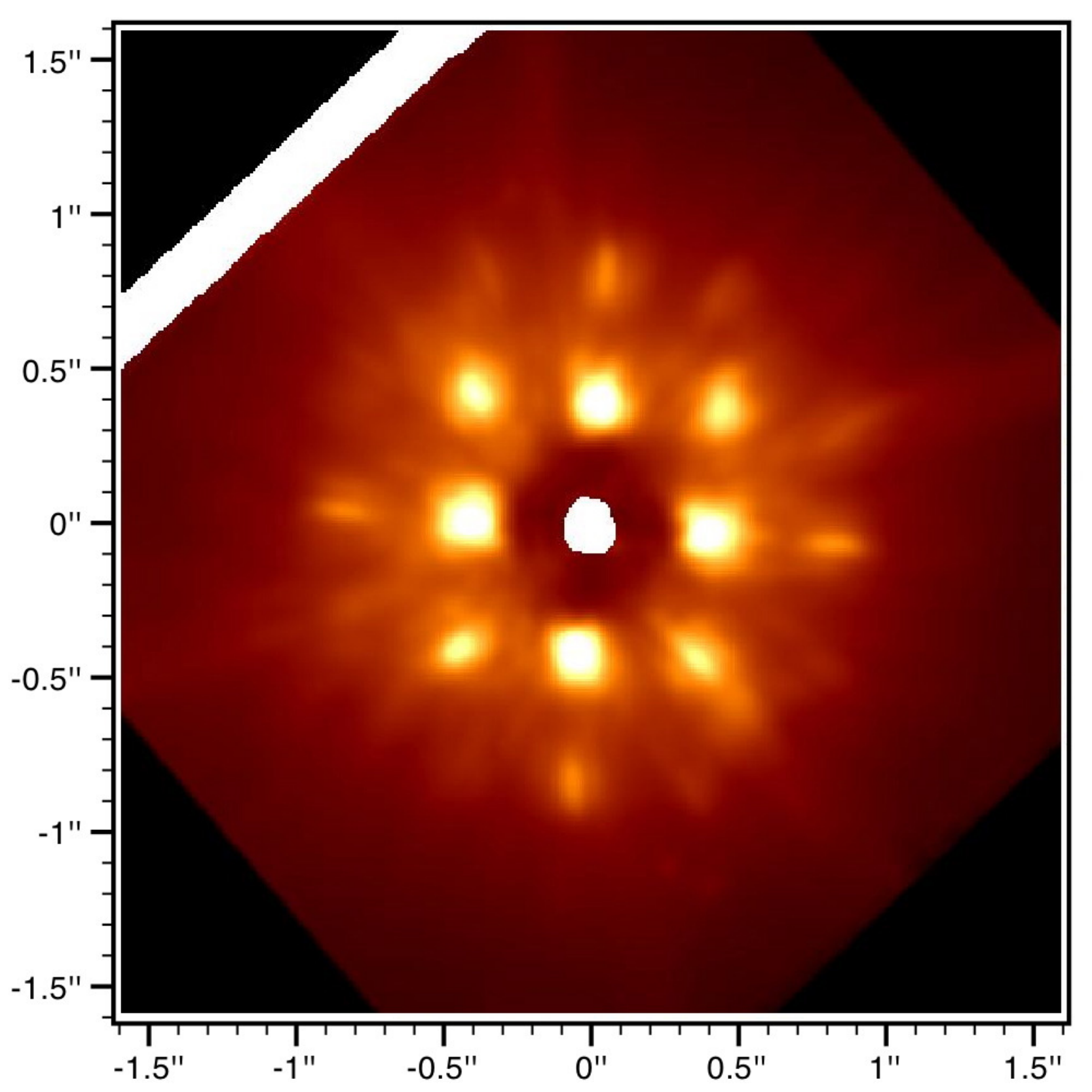} \\
\includegraphics[width=4.4cm]{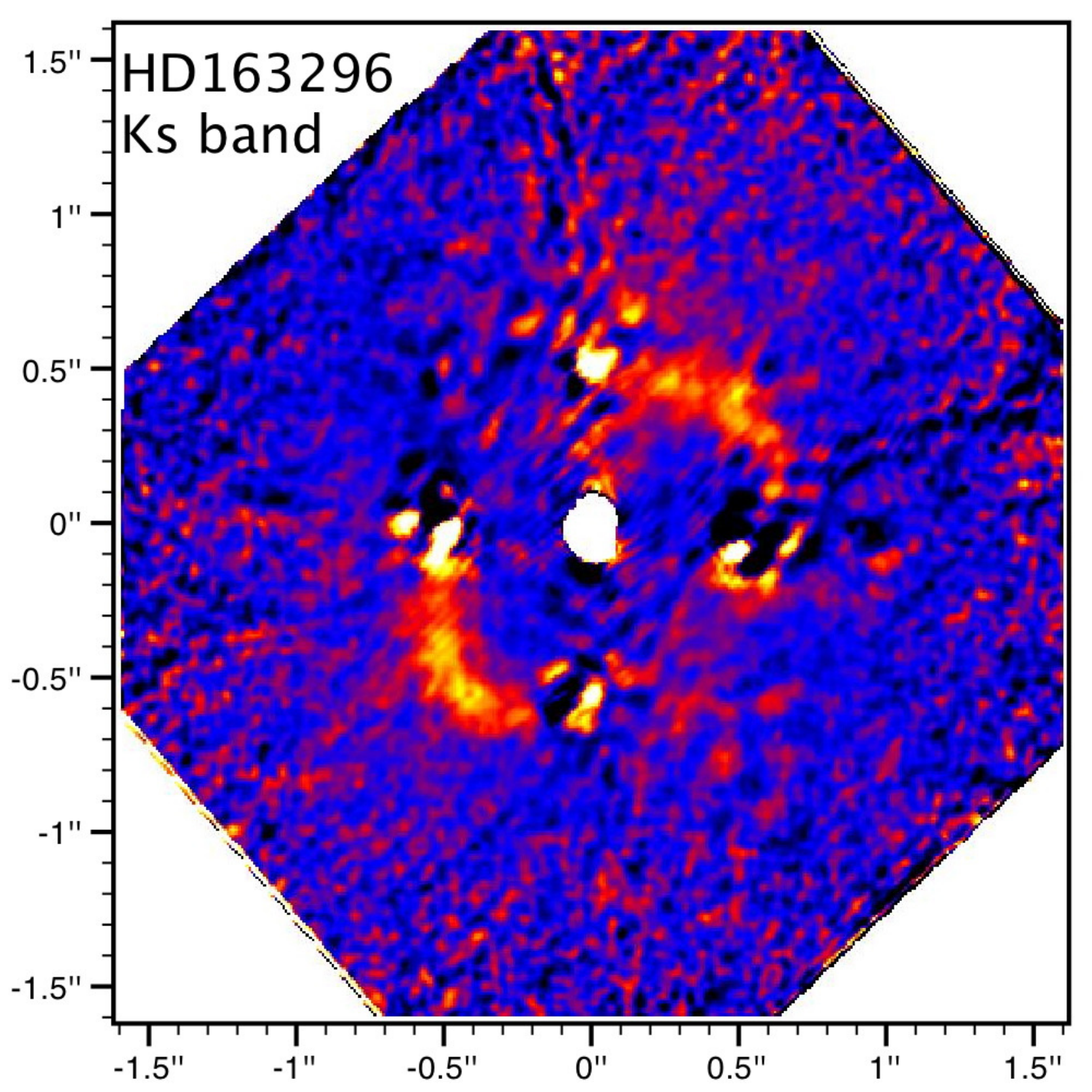}
\includegraphics[width=4.4cm]{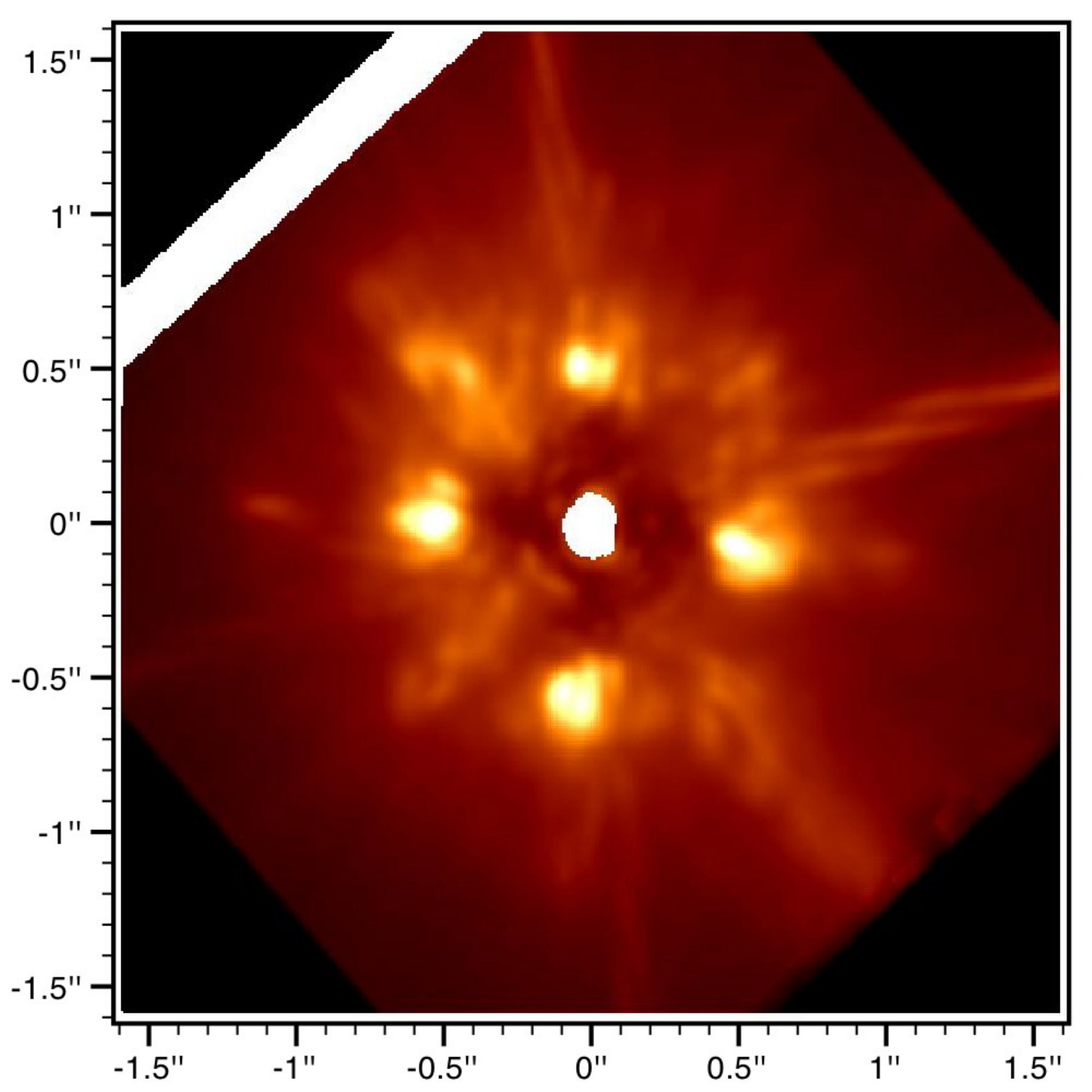} \\
 \includegraphics[width=4.4cm]{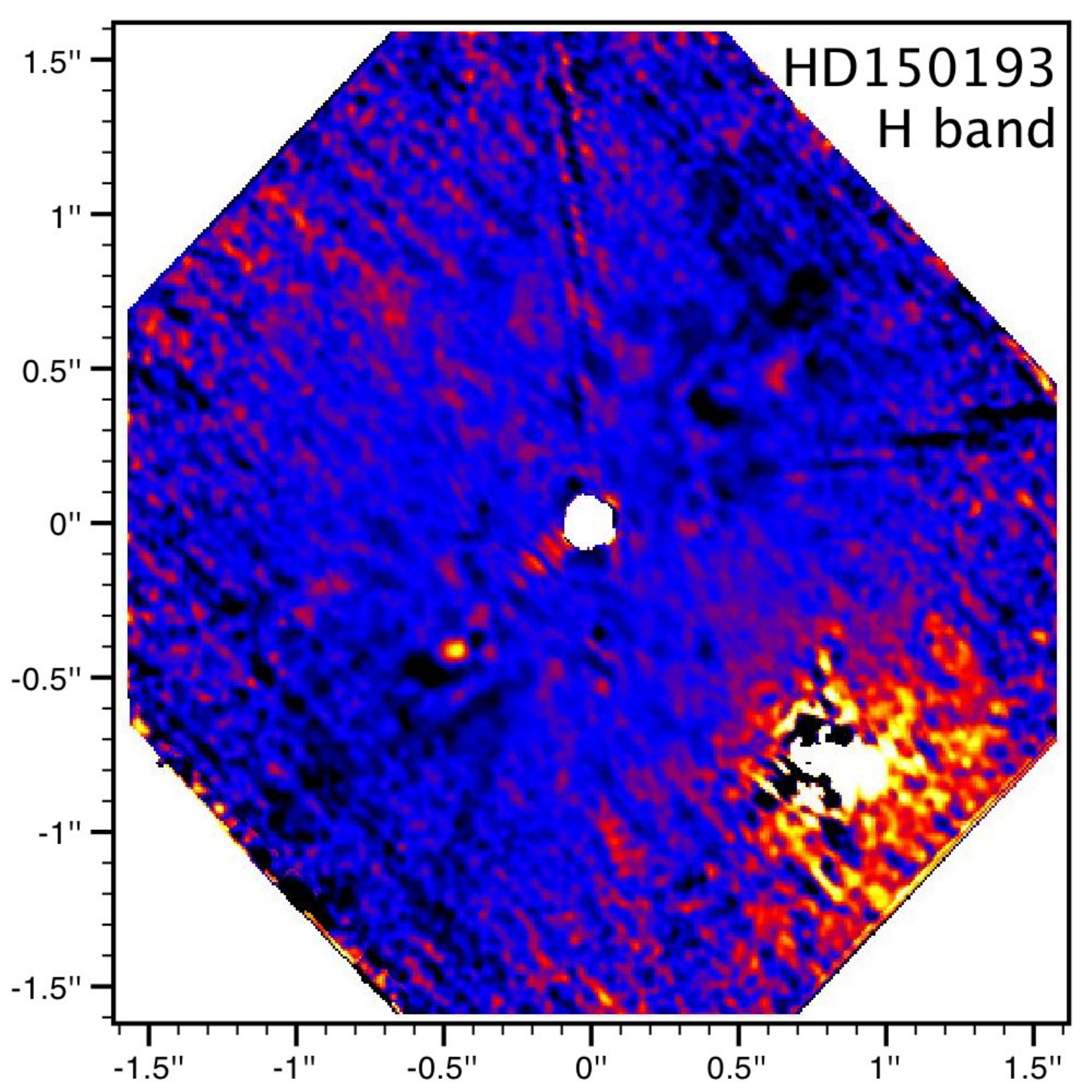}
\includegraphics[width=4.4cm]{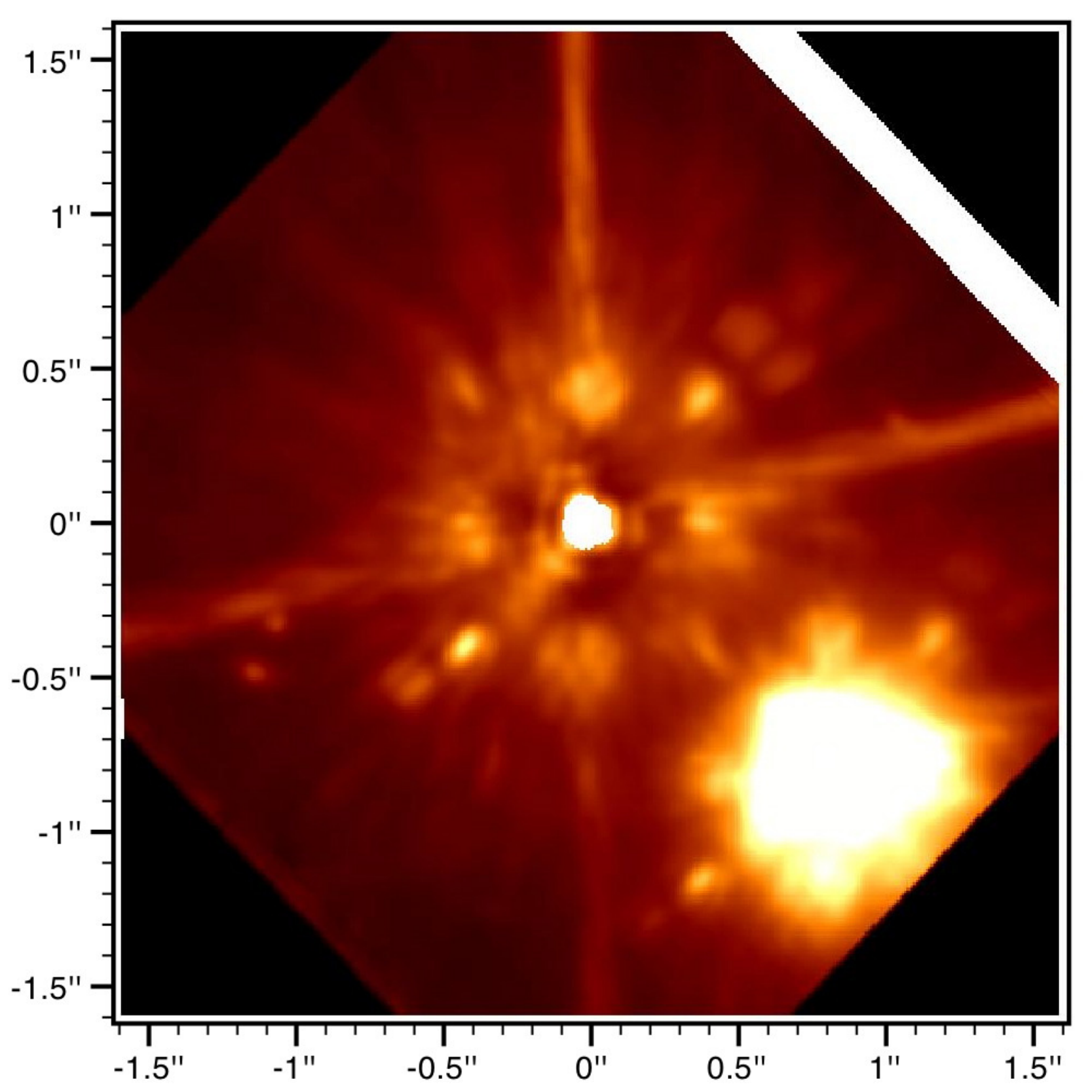} \\
\includegraphics[width=4.4cm]{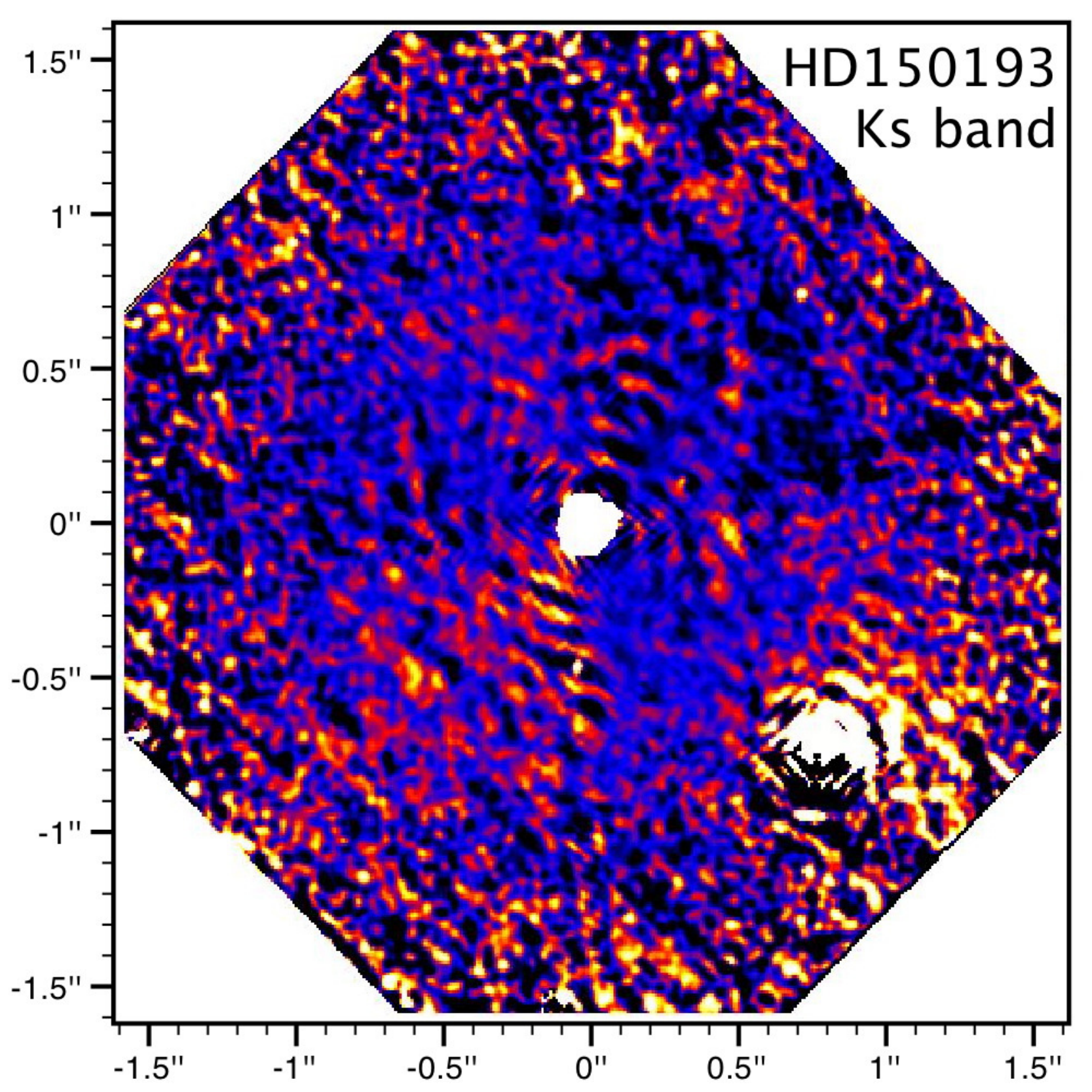}
\includegraphics[width=4.4cm]{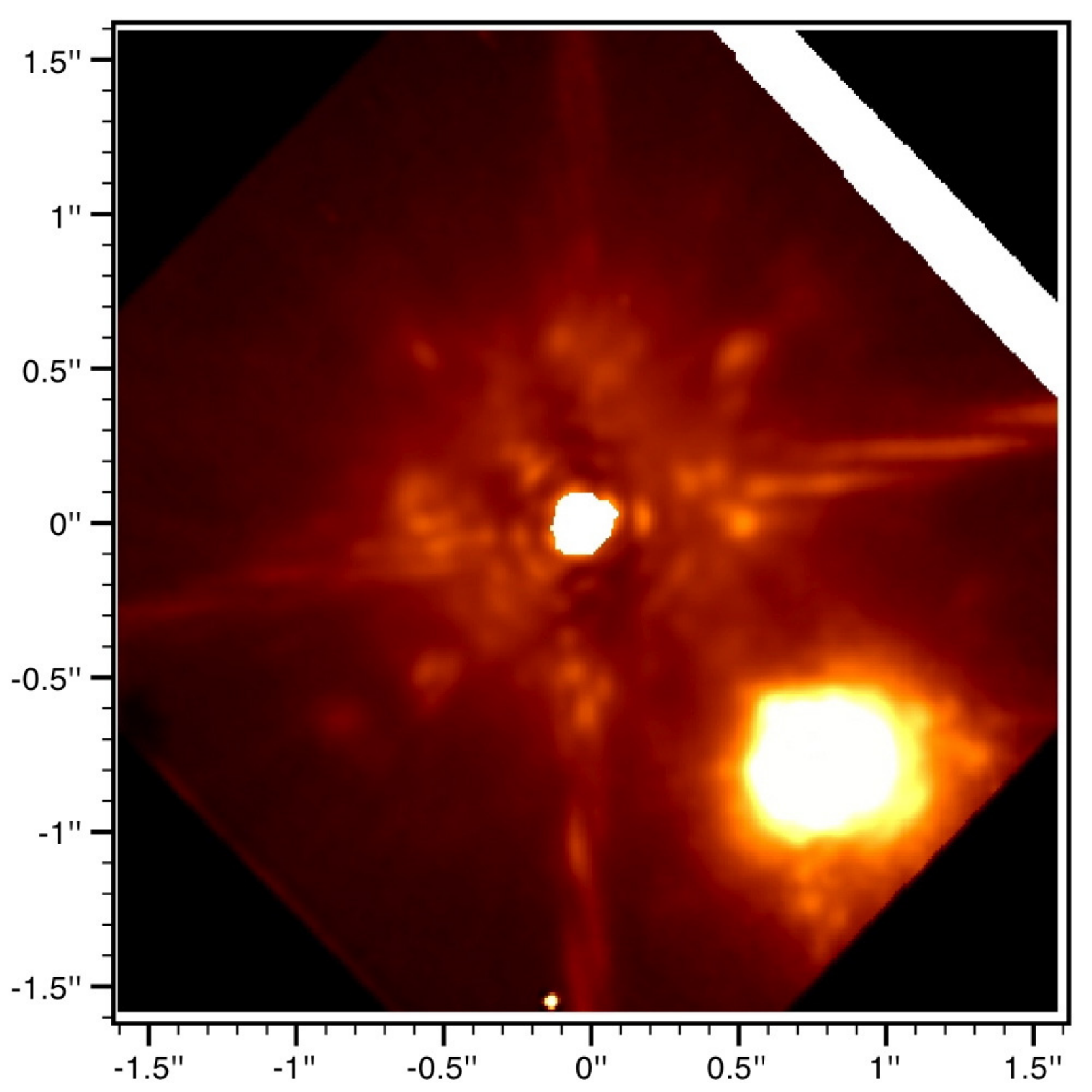} \\
 \includegraphics[width=4.4cm]{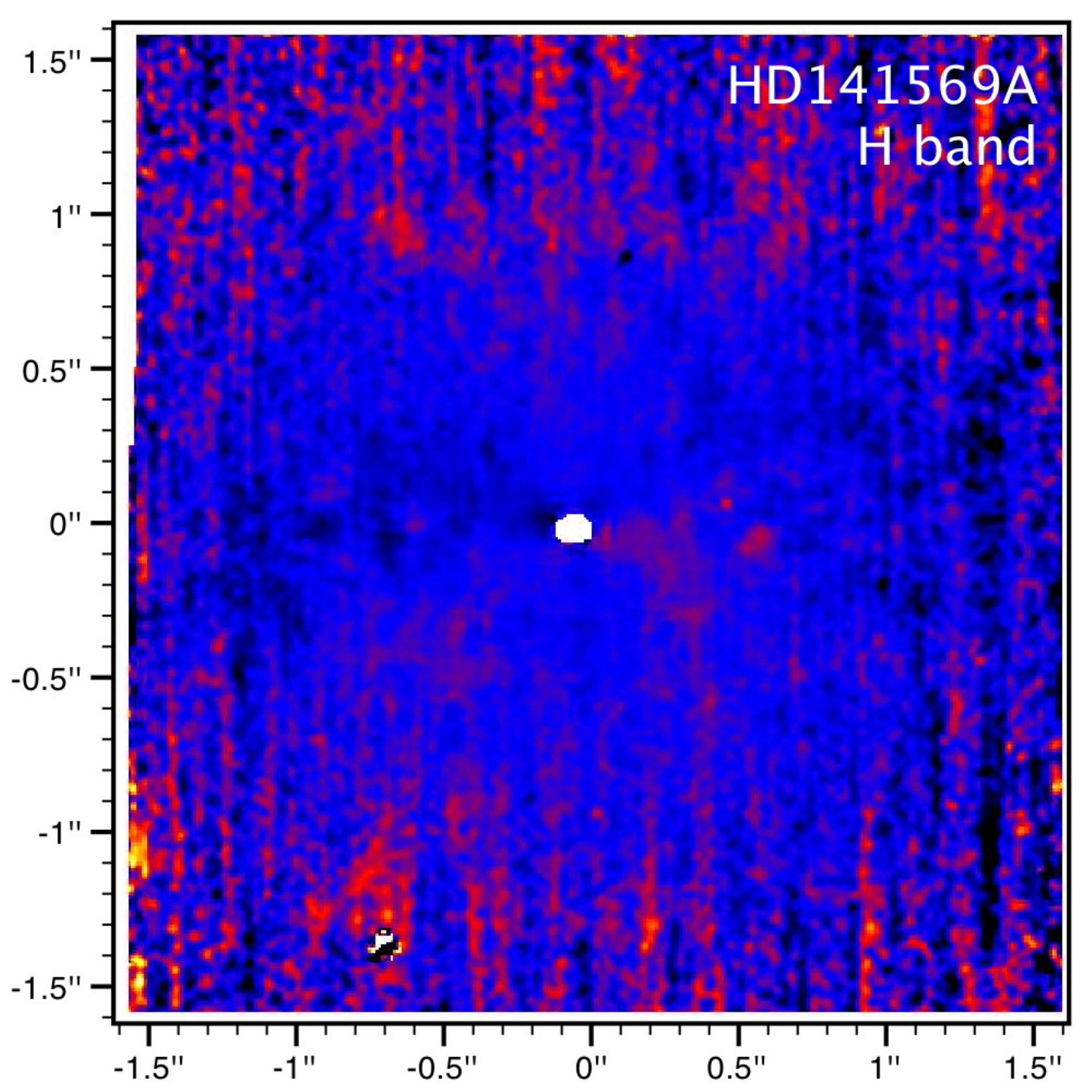}
\includegraphics[width=4.4cm]{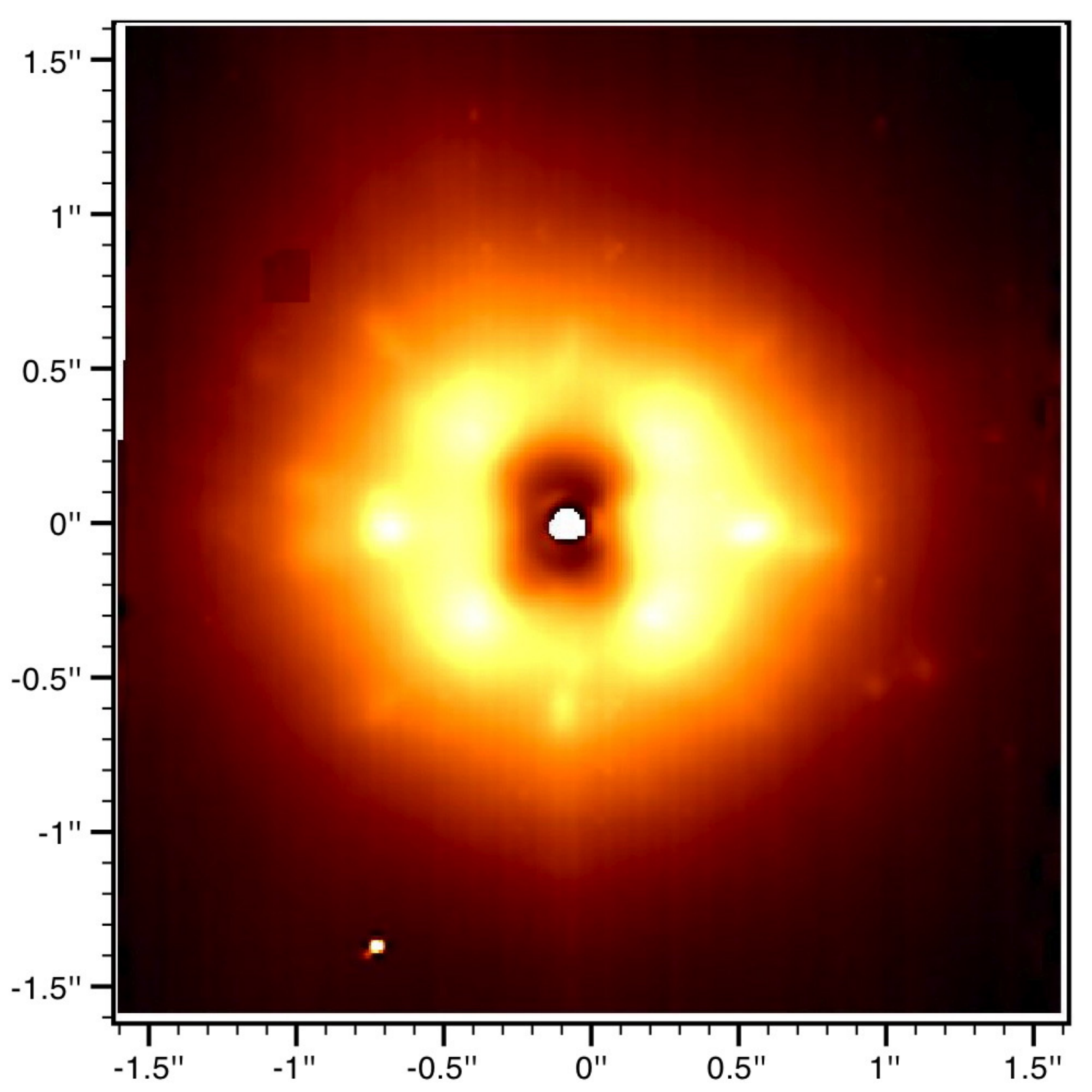}
  \caption{Polarized-light $Q_{\rm T}$ images (left column) and intensity $I$ images (right column) of HD163296, HD150193, and HD141569A. North is up, east is left. The white central area denotes the saturated pixels, which are not included in the analysis. All images are scaled by $r^2$ to compensate for stellar dilution. This enhances the brightness of the companion and of all outer speckles. All $Q_{\rm T}$ images are shown with the same linear scale, whereas the scale of the $I$ images is arbitrary to highlight each PSF.}
            \label{Images}
  \end{figure}

In our analysis, we used the tangential Stokes parameters $Q_{\rm T}$ and $U_{\rm T}$, defined as
\begin{equation}
\begin{split}
Q_{\rm T} = +Q \cos(2\phi) + U \sin(2\phi) \\
U_{\rm T} = -Q \sin(2\phi) + U \cos(2\phi)
\end{split}
,\end{equation}
with $\phi$ being the polar angle of a given position with respect to a polar coordinate system centered on the star, and $Q$ and $U$ are the Stokes parameters measured at half-wave-plate position angles of 0$^\circ$/-45$^\circ$ and -22.5$^\circ$/-67.5$^\circ$ , respectively \citep[see, e.g.,][]{Schmid2006}. In systems with only tangential polarization, $Q_{\rm T}$ is by construction equal to $P=\sqrt{Q^2+U^2}$ but does not generate a biased noise because of the square calculation. On the other hand, $U_{\rm T}$ should not contain any signal and can be used to estimate the error. Here, errors were calculated from the standard deviation of pixel values in a resolution element of both the $Q_{\rm T}$ and $U_{\rm T}$ images. Finally, the intensity $I$ images of the sources were obtained by summing the contributions from ordinary and extraordinary beams of each image and then averaging them. The $I$ images are useful for determining any potential artifacts in the final $Q_{\rm T}$ frames that result from imperfect PSF subtraction.

\section{PDI images} \label{PDI_results}

In Fig.\,\ref{Images} we show the PDI images that result from the data reduction described above.  The $I$ images are shown in the right column. These images differ significantly from each other, and numerous artifacts are visible. The $Q_{\rm T}$ images are shown in the left column. Only the disk around HD163296 is detected. Therefore, we dedicate Sect.\,\ref{Images_HD163296} to this object alone, whereas the two non-detections are described in Sect.\,\ref{Non-detections}.

\subsection{HD163296} \label{Images_HD163296}
The $H$ and $K_{\rm S}$ band PDI images of HD163296 are shown in the first two rows of Fig.\,\ref{Images}. All images are contaminated by AO residuals, as seen in the intensity images in the right column. Features at these locations must be considered spurious. However, the $Q_{\rm T}$ images of HD163296 show an extended ring structure at $\sim 0.6\arcsec$ from the star with no counterpart in the $I$ images. This structure is much more evident in the $K_{\rm S}$ band, but it is revealed in the $H$ -band image as well. Note, however, that there are eight AO spots in the $H$ image compared with only four in $K_{\rm S,}$ and they are generally brighter. This may affect any potential weak detections from the $Q_{\rm T}$ images in the $H$ band.  

\begin{figure*}
   \centering
 \includegraphics[width=15cm]{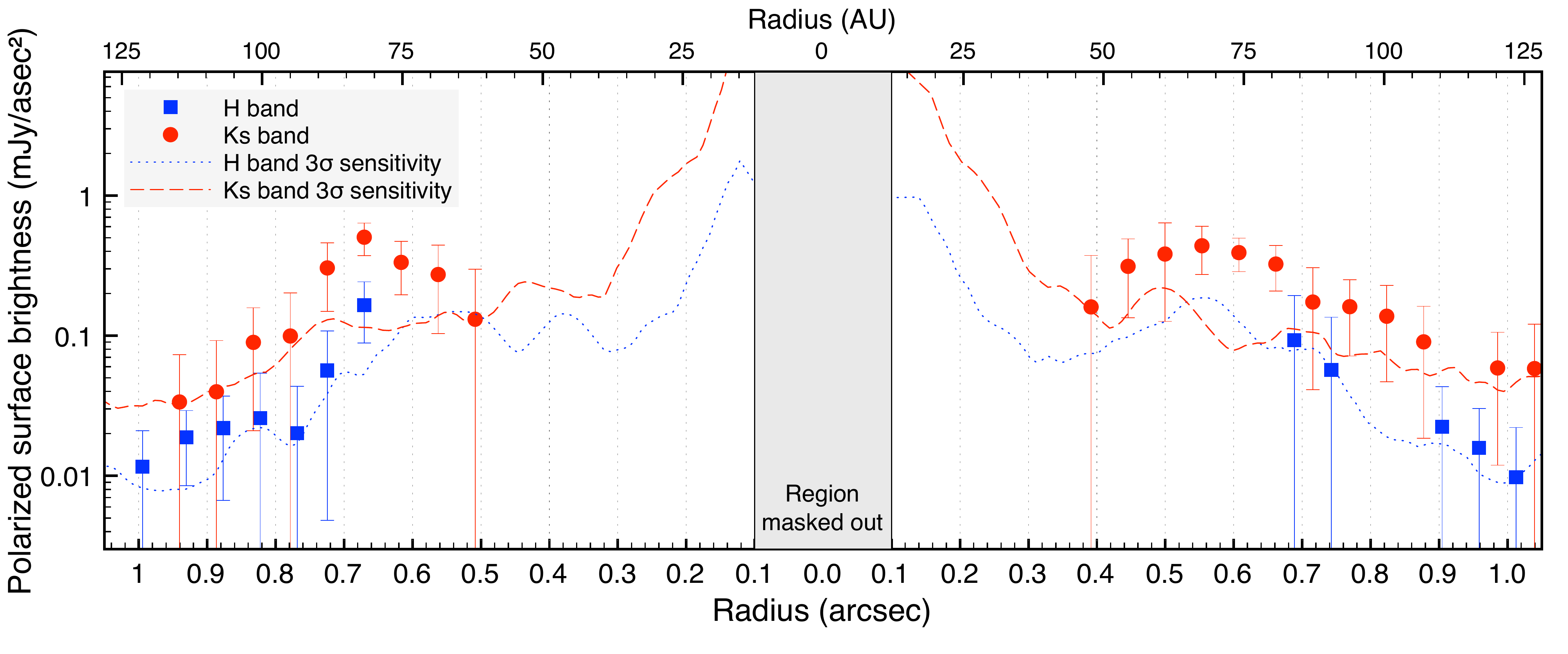}
 \includegraphics[width=15cm]{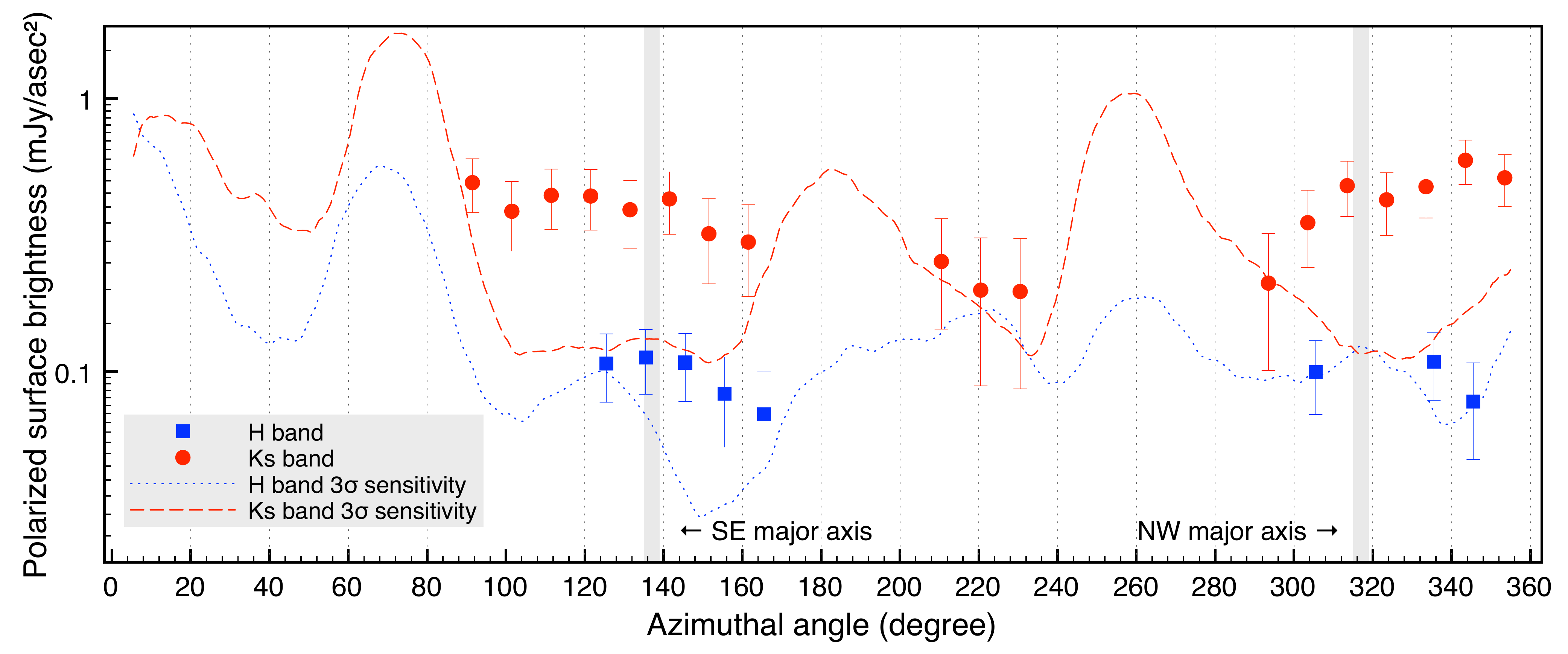}
  \caption{Polarized emission distribution from HD163296 in $H$ and $K_{\rm S}$ band. Top: radial profile obtained from a 3 pixel-wide cut along the major axis (left side is East). Bottom: azimuthal profile obtained integrating over concentric ellipses from $0.6\arcsec$ to $0.7\arcsec$ (see text for details). Only detections with more than $3\sigma$ are shown. The errors and the sensitivity are calculated from the local noise in both the $Q_{\rm T}$ and $U_{\rm T}$ images. The sensitivity bumps are due to the AO spots. Systematic errors from the photometric calibration are not included. The vertical grey stripes in the azimuthal profile indicate the location of the major axis.}
            \label{Radial}
  \end{figure*}

\textit{Radial profile.} The ring feature can be clearly traced to the NW and SE. We also report a tentative detection  to the SW (only in the $K_{\rm S}$ ). The ring pattern is interrupted by the southern and western AO spots. The radial width of this feature is always larger than the angular resolution of the observations ($>0.09\arcsec$). This suggests that the structure is radially resolved. The brightest regions of the ring lie along the disk major axis \citep[PA=137$^\circ$, from submm images by][]{deGregorio-Monsalvo2013}. Therefore, we made a three-pixel wide cut along this direction both in the $H$ and $K_{\rm S}$ image and plot the radial profile (see top panel of Fig.\,\ref{Radial}). Emission above the $3\sigma$-level is detected in the $K_{\rm S}$ band from $0.5\arcsec$ to $0.95\arcsec$ and from $0.4\arcsec$ to $1.0\arcsec$ (east and west side). In
the $H$ band the emission is revealed from $0.6\arcsec$ to $1.0\arcsec$ along both sides. The polarized surface brightness varies from 0.5 to 0.03 mJy/arcsec$^2$ in the $K_{\rm S}$ band and from 0.15 to 0.01 mJy/arcsec$^2$ in the $H$ band. The inner edge of the ring structure does not appear  to be sharp. In fact, the polarized-light distribution decreases smoothly from the peak luminosity to the noise level over $\sim 0.2\arcsec$.

\textit{Azimuthal profile.} Along the SW minor axis, we detect a tentative emission in the $K_{\rm S}$ band (from 0.4$\arcsec$ to 0.5$\arcsec$). Given the radial profile along the major axis and assuming a circular disk, this translates into an inclination $i=43^{\circ \,+9^\circ}_{-14^\circ}$. This value is consistent with what was inferred by \citet{deGregorio-Monsalvo2013} with submm imaging ($i=45^{\circ}$). To investigate the azimuthal distribution of the structure, we fit concentric ellipses on a $45^{\circ}$ plane. The ellipses match the ring maxima better by introducing a marginal offset along the north-south ($0.05\arcsec$) and east-west ($0.03\arcsec$) directions. Finally, we integrated over concentric ellipses from 0.6$\arcsec$ to 0.7$\arcsec$. The profile thus obtained (see bottom panel of Fig.\,\ref{Radial}) suggests that the emission in $K_{\rm S}$ band is maximized along a wedge around the major axis ($\sim137^{\circ}$ and $\sim317^{\circ}$). The average brightness ratio between the major and minor SW axis is $\sim 2$. The only contribution in the $H$ band is also detected on the major axis. However, the sensitivity anywhere else is worse than along the major axis (because of the AO spots), and the contrast between the major and minor axis cannot be constrained.

\textit{[$H-K_S$] color.} Since we detected emission along the East major axis from both bands, we calculated the local [$H-K_{\rm S}$] color and compared it with the stellar one. To do this, we considered an average emission from a three-pixel wide cut from $0.6\arcsec$ to $0.7\arcsec$ and used the 2MASS stellar color for the source \citep{Cutri2003}.
The local disk [$H-K_{\rm S}$] color is $1.92^{+1.29}_{-1.19}$, whereas the star shows a value of $0.75^{+0.08}_{-0.04}$. The disk emission is therefore likely to be redder than the stellar one (these magnitudes translate into a flux ratio of roughly three). However, the small overlap of the error bars (dominated by the photometric calibration uncertainty of $\sim40\%$) does not allow us to draw definitive conclusions.

\subsection{HD141569A and HD150193A} \label{Non-detections}
The final images of HD141569A and HD150193A do not reveal any extended emission in polarized light. As shown in the third to fifth row of Fig.\,\ref{Images}, any feature in the $Q_{\rm T}$ images corresponds to an artifact in the $I$ images and is thus ascribable to an imperfect PSF subtraction. 

In particular, the intensity images of HD150193A reveal the K4 companion at 1.11$\arcsec \pm 0.03\arcsec$. This separation is consistent within the error bars with what was obtained one decade before by \citet[1.10$\arcsec \pm 0.03\arcsec$]{Fukagawa2003}. The $Q_{\rm T}$ images are brighter around HD150193B because of the $r^2$ intensity scaling, but the signal-to-noise ratio does not diverge from a stochastic distribution. 

\section{Discussion: the broken ring of HD163296} \label{Discussion_HD163296}
Since the $Q_{\rm T}$ images are expected to contain any polarized fraction of light that is tangentially scattered, there is little doubt that the ring structure described in Sect.\,\ref{Images_HD163296} is tracing (part of) the circumstellar disk that is directly illuminated by the central star. The distribution of polarized scattered light is strongly affected by the disk geometry and the dust properties. Therefore, the relation between these PDI maps and the actual dust distribution is not straightforward. In this section, we discuss the effects that may cause this "broken ring" structure.

\subsection{\textit{Why broken?}} \label{Discussion_broken}
The strongly azimuthally asymmetric structure revealed by our PDI images is the result of both the anisotropic scattering function (hereafter $f_{\rm scat}(\theta)$) and the scattering-angle dependence of the polarization fraction ($f_{\rm pol}(\theta)$). Since no scattered-light image of this object has reached such small inner working angles and the disk-to-star contrast is not high enough for the disk to be detected from our intensity images, we cannot distinguish these two effects. However, scattered-light images of the outer disk \citep{Grady2000}, spectroscopic observations of the jet \citep{Ellerbroek2014}, {and the spatial morphology of the CO emission lines \citep{Rosenfeld2013}} suggest that the NE region of the disk is the near side.

To constrain $f_{\rm scat}(\theta)$ and $f_{\rm pol}(\theta)$, we first obtained a rough estimate on the actual scattering angles, assuming that only the disk surface scatters light. The scattering angle $\theta$ for light from the major axis can be roughly estimated through $\theta=90^{\circ}-\beta$, with $\beta$ being the disk opening angle. According to the gas vertical structure of this disk proposed by \citet{deGregorio-Monsalvo2013} and \citet{Rosenfeld2013}, $\beta$ is $\sim 20^{\circ}$ at 80 AU, and therefore $\theta\simeq70^{\circ}$. The scattering angle along the minor axis is given by $\theta=90^{\circ}\pm i-\beta$, where $i$ is the disk inclination. The plus sign is for the far side, yielding $\theta\simeq115^{\circ}$, and the minus sign for the near side, yielding $\theta\simeq25^{\circ}$.

From the $Q_{\rm T}$ images of Fig.\,\ref{Images} it is clear that the brightness distribution is maximized along the direction of the major axis. In particular, the polarized emission from the major axis ($\theta\simeq70^{\circ}$) is twice as high as that from the SW minor axis (the far side, $\theta\simeq115^{\circ}$, detected in the $K_{\rm S}$ band) and \textit{at least} 1.5 times higher than that from the NE minor axis (the near side, $\theta\simeq25^{\circ}$, undetected, but constrained from the $3\sigma$-sensitivity). The scattering function due to interstellar dust monotonically decreases with the angle, meaning that $f_{\rm scat}(25^{\circ})>f_{\rm scat}(70^{\circ})>f_{\rm scat}(115^{\circ})$. Therefore, the observed azimuthal distribution of polarized light translates into limits on the polarization fraction distribution: 
\begin{equation}
\begin{split}
f_{\rm pol}(70^{\circ})\gtrsim 1.5\cdot f_{\rm pol}(25^{\circ})\ \ \\ 
f_{\rm pol}(70^{\circ})\lesssim 2.0\cdot f_{\rm pol}(115^{\circ})
\end{split}
.\end{equation}
Given the typical $f_{\rm pol}(\theta)$ curves \citep[e.g.,][]{Murakawa2010, Min2012}, these constraints are satisfied by all dust grain sizes and types. In other words, flared and moderately inclined disks might show a polarized-light distribution maximized along the major axis, similarly to HD163296, regardless of their grain properties.

The forward-scattering nature of dust may also suggest that the detection from the far side only is a contradiction. We note, however, that the images are noisier along the near side (see Fig.\,\ref{Radial}). Given our sensitivity, we can only claim that the near side is not much brighter than the far side (less than a factor 1.5). This is not necessarily suggestive of an isotropic scattering function, because the high polarizing efficiency at $115^{\circ}$ can compensate for a lower $f_{\rm scat}$ at those angles \citep[like in HD142527,][]{Avenhaus2014}. To constrain the scattering asymmetry, an assumption on the $f_{\rm pol}(115^{\circ})/f_{\rm pol}(25^{\circ})$ ratio is thus needed. By looking at the typical $f_{\rm pol}(\theta)$ curves at $\lambda\simeq2\,\mu$m \citep{Perrin2009, Murakawa2010}, this ratio varies with dust grain size and composition from 2.5 to 12. Assuming a conservative value of 12, the above claim yields
\begin{equation}
f_{\rm scat}(25^{\circ})\lesssim 18 \cdot f_{\rm scat}(115^{\circ})
.\end{equation} 

This constrain in turn translates into a Henyey-Greenstein asymmetry parameter $g\lesssim0.6$. The parameter thus obtained is only an approximation of the real $g$ \citep{Henyey1941}, which can be obtained even though the knowledge of $f_{\rm scat}(\theta)$ is limited to a few values of $\theta$. In near-IR, values lower than 0.6 are expected for sub-$\mu$m size particles \citep{Pinte2008}. However, particles can potentially be so forward-scattering that most of the radiation is scattered by $\theta<25^{\circ}$. In that case, the approximated $g$ significantly diverge from the real asymmetry parameter \citep[e.g.,][]{Mulders2013}.

Finally, the red color inferred in Sect.\,\ref{Images_HD163296} can also provide insight into the grain size. An enhanced flux in the $K_{\rm S}$ band with respect to the $H$ band (roughly three times higher than the stellar ratio) can be either explained with a dust albedo much higher at wavelengths $\gg$ 2.2 $\mu$m (as for larger $\mu$m-sized particles) or with a high difference between the polarizing efficiency for $H$ and $K_{\rm S}$ band \citep[as for sub-$\mu$m sized particles, see Fig.\,1 of][]{Murakawa2010}. In particular, the latter hypothesis is consistent with what is suggested by the brightness asymmetry. Alternatively, red colors from the grains may be also provided by particular disk geometries (see Sect.\,\ref{Discussion_self-shadowing}).

Summarizing, the maximization of the polarized flux along the major axis is not surprising and might be expected for all flared and inclined disks. On the other hand, the non-detection of the disk near side and the overall red color of the disk may suggest very small particles on the disk surface. However, deeper observations with a smaller photometric uncertainty are needed to draw significant conclusions on the dust population. 

\subsection{\textit{Why a ring?}} \label{Discussion_ring}
Along the SE-NW axis, the $Q_{\rm T}$ images depict the disk of HD163296 as a ring structure. However, the observed inner and outer edge of the polarized-light emission may not represent the physical dust distribution. In particular, the outer edge at $\sim 1\arcsec$ is set by the sensitivity of these observations (see radial profile of Fig.\,\ref{Radial}). Disks observed in PDI are often truncated at $\sim1.5\arcsec$ \citep[e.g.,][]{Quanz2011, Quanz2013}. The fact that the OWA of these observations is slightly smaller is probably connected to the intrinsically weaker scattered emission from the disk.

The interpretation of the inner edge is instead less univocal. Schematically, a depletion of polarized emission from the disk inner regions can be due to four factors: ($i$) a lack of photons that reach the disk surface (self-shadowing), ($ii$) a substantial alteration of disk surface geometry, ($iii$) an intrinsic deficit of scattering particles, ($iv$) a dramatic change in the dust properties. We now discuss the possible reasons for the observed light drop in more detail.

\subsubsection{Self-shadowing} \label{Discussion_self-shadowing}
If the innermost part of the disk is puffed up, it can cast a shadow on the outer part of the disk. However, if the disk is flared, its surface will re-emerge from the shadow at a certain distance from the star \citep[e.g.,][]{Dullemond2001}. Therefore, self-shadowing is more likely to occur in flat (group II) than in flared disks (group I). This effect has been proposed to explain the diverse shapes of the far-IR SED typically observed in disks, but it can also be applied to PDI images to explain the detection of annular gaps \citep[e.g.,][]{Quanz2013}. 

For HD163296, a variable self-shadowing has been proposed by \citet{Wisniewski2008} to explain the variable scattered-light intensity observed in the $>2\arcsec$ outer disk. This scenario is supported by the variable scale height of the inner disk wall, as inferred from near-IR SED monitoring \citep{Sitko2008}. Interestingly, this source is often referred to as a border-line object between a group I and II \citep[e.g.,][]{Meijer2008}. Thus, the disk geometry may be such that only a fraction of the outer disk surface is shaded. At the time of our observations the corresponding radius would be $\sim70$ AU, but it might change over time, depending on the exact geometry of the inner wall. This morphology may also explain why the detection of the ring in the $H$ band is weaker. Radiation is subject to higher extinction at 1.6 $\mu$m than at 2.2 $\mu$m and some extent of the inner wall may allow the passage of light in a wavelength-dependent manner.

In this scenario, second-epoch PDI observations might detect the inner edge of the emission at a different location. Furthermore, observations shortward of the $H$ band might detect an even weaker emission from the disk.

\subsubsection{Flaring angle}
Even in the absence of self-shadowing, an abrupt change of the disk scale height may still explain our observations. This is because the amount of scattered light is strongly dependent on the incident angle. To estimate the disk scale height enhancement necessary to cause the observed "jump" in polarized light, we used the difference between the disk-to-star light contrast inside and at the location of the ring (a factor $\sim6$, Sect.\,\ref{Discussion_dust}). We considered the phase functions and polarization degree curves by \citet{Min2012} and focused on the $45^{\circ}-90^{\circ}$ interval (reasonable scattering angles along the major axis). In this range of scattering angles, the trend of the two curves is opposite (the scattering function decreases, the polarization efficiency increases). Thus, the net effect is limited. In particular, a contrast increasing by a factor 6 can only be explained with a disk opening angle that increases by more than $20^{\circ}$ over $\sim 40$ AU. This translates into a disk flaring index\footnote{$\beta$, as from $H(r)/H_0=(r/r_0)\,^{\beta}$} of 1.9, which is considerably higher than the typical strong flaring values \citep[$\sim1.2$, e.g.][]{Woitke2010}. Thus, a "normally" flared disk cannot explain these images. However, a particularly puffed-up geometry beyond $\sim 50$ AU is still a valid option. Interestingly, \citet{Dominik2003} showed that the mid-IR spectrum of this source behaves as those of flat disks, whereas an uncommon bump at 100 $\mu$m cannot be fitted by a continuous flaring disk.

\subsubsection{Depletion of dust}
A relevant depletion of $\mu$m- to mm-size grains in the inner few tens of AU has been observed in many circumstellar disks (the so-called transition disks). This morphology is reflected in a deficit of flux from near- to mid-IR wavelengths \citep{Strom1989}. At first glance, the morphology of the disk of HD163296 may resemble those of some transition disks observed in PDI \citep[e.g.,][]{Quanz2011, Hashimoto2012}. However, the smooth radial profile of Fig.\,\ref{Radial} suggests that the apparent disk inner edge at $r\simeq0.6\arcsec$ does not intercept a large fraction of radiation (i.e., it is not a disk wall). Furthermore, we are not aware of any complementary dataset pointing toward the presence of a cavity, either from SED \citep{Isella2007} or from submm imaging \citep{deGregorio-Monsalvo2013}.

\subsubsection{Dust properties} \label{Discussion_dust}
As discussed in Sect.\,\ref{Discussion_broken}, the polarized surface brightness is very sensitive to the dust properties. Macroscopic changes therein can effectively modify their albedo and/or polarizing efficiency and, thus, provide a substantially different polarized intensity.

One of the key processes is grain growth. This process causes a significant decrease in the optical depth and has therefore been suggested as a possible explanation for the large cavities of transition disks \citep[e.g.,][]{Dullemond2005}. Typically, this theory fails to explain the observed sharp change in the radial surface brightness and the rapid creation of the large cavities observed at (sub-)mm wavelengths {\citep{Birnstiel2012}}. However, the case of HD163296 does not require an explanation
of these two features, since no cavity is observed at submm wavelengths \citep{deGregorio-Monsalvo2013} and the radial light drop at the inner edge is indeed smooth (see Fig.\,\ref{Radial}). In particular, the growth of scattering grains acts to decrease the observed polarized flux by three means: by lowering the intrinsic dust albedo, by providing a more forward-peaking phase curve (and thus a deficit at scattering angles larger than $20^{\circ}$), and by flattening the polarizing efficiency curve. Therefore, small variations in the grain size may be sufficient to explain the estimated contrast drop ($\simeq6$) and thus the observed disk morphology. A more quantitative approach would require an ad-hoc radiative transfer model that is beyond the scope of this paper. 

\begin{figure}
   \centering
 \includegraphics[width=8cm]{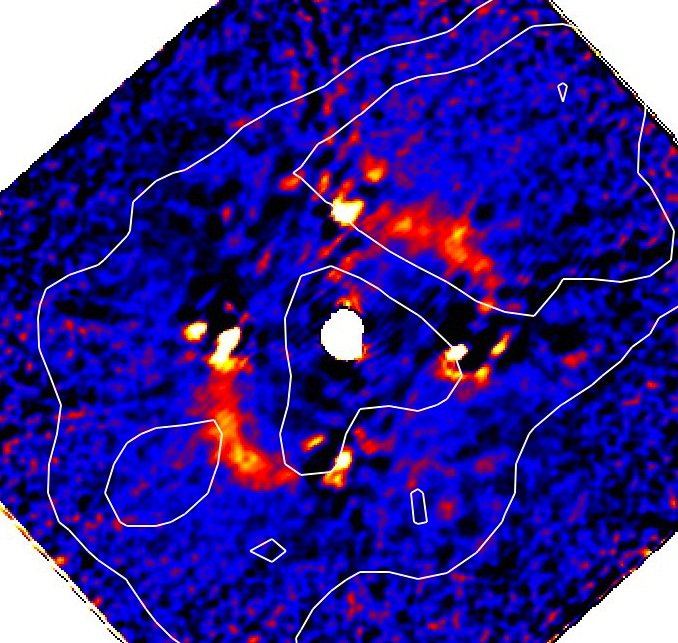}
 \includegraphics[width=9cm]{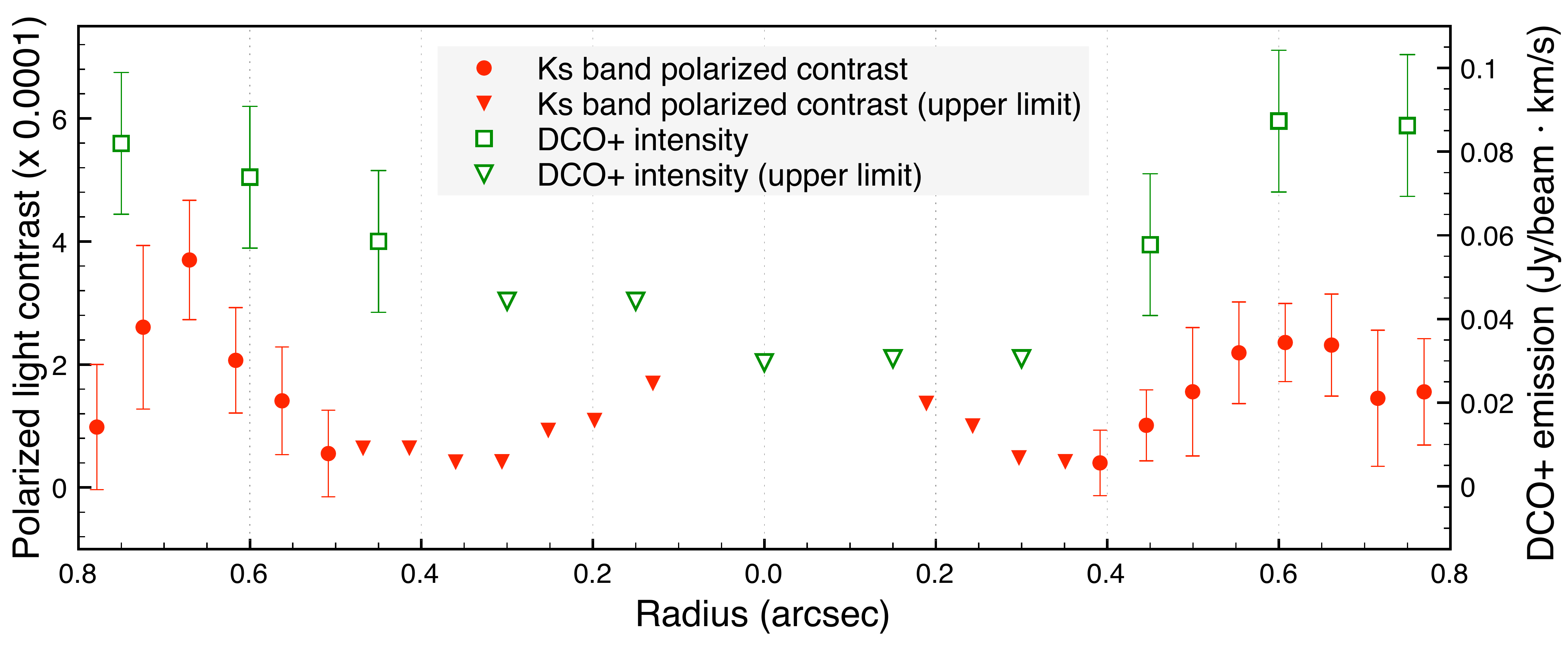}  
  \caption{Top: DCO+ map from the disk of HD163296 \citep{Mathews2013} superimposed on our PDI $K_{\rm S}$ -band image. The white contours indicate the emission at 3$\sigma$ and 4$\sigma$ level ($\sigma=18 {\rm mJy\ beam^{-1}\ km\ s^{-1}}$). The DCO+ emission is depleted inside $\sim0.5\arcsec$ , while it reaches its maximum roughly at the location of the dusty ring. Bottom: DCO+ intensity along the major axis vs contrast of the PDI images calculated from Eq.\,\ref{Formula_albedo}.}
            \label{DCO+}
  \end{figure}

An alternative explanation invokes the role of ice in the disk structure. Dust grains are expected to have a much higher albedo when coated with icy molecules on their surface \citep[e.g.,][]{Inoue2008}. Molecules of a certain species are predicted to freeze out below temperatures typically found in the disk mid-plane (i.e.,\ beyond the so-called snow line). HD163296 is one of the few objects for which the location of the CO snow line has been inferred \citep{Qi2011}. More specifically, the location where the CO begins to freeze out can be traced with the inner edge of the expected ring emission of DCO+ \citep{Mathews2013}. In Fig.\,\ref{DCO+}, we show a comparison of the DCO+ emission map from HD163296 \citep[ALMA Science verification data,][]{Mathews2013} with our PDI image. We use the disk-to-star contrast at a certain radius $r$ defined as
\begin{equation} \label{Formula_albedo}
\phi_{\rm pol}=  \frac{F_{\rm pol}}{F_*/4\pi r^2} 
,\end{equation}
with $F_*$ being the stellar flux in $K_{\rm S}$ band and $F_{\rm pol}$ the polarized flux measured along the major axis of our images. Note that $\phi_{\rm pol}$ indicates the polarized fraction of the scattered-light contrast. This is, in turn, a combination of both the dust properties and the disk geometry and can therefore substantially differ from the intrinsic dust albedo \citep[particularly if the scattering function is highly anisotropic,][]{Mulders2013}. As we see from Fig.\,\ref{DCO+}, we obtain a factor $\Delta \phi_{\rm pol}\gtrsim 6$ difference between $0.65\arcsec$ (the location of the peak intensity) and $0.3\arcsec$ (representative radius inside the non-detection region). Interestingly, the transition from the noise level to the peak intensity of DCO+  that roughly occurs over the same interval. This may suggest that in this region mantles of CO on the grain surface are becoming thicker and can therefore increase the dust albedo by the $\Delta \phi_{\rm pol}\gtrsim 6$ necessary for the emission from our dataset to rise over the sensitivity. 

Although it is tantalizing, the ice explanation presents a number of caveats. The first, instrumental, is related to the low angular resolution of the ALMA data ($0.65\arcsec \times 0.44\arcsec$) that prevents us from drawing strong conclusions from the possible spatial coincidence of the two datasets. The second is geometrical; since our data trace the disk surface, whereas the DCO+ emission is thought to originate from a deeper region in the disk mid-plane, one may expect the location of the CO snow line to be not coincident and, in particular, to be located farther out when traced by PDI images. A final point is that ice mantles on the grain surface may decrease the polarizing efficiency of grains because of the
multiple scattering therein. Therefore, the effects on the albedo and on the polarization are competing and, for this theory to explain our data, the former must be the dominant.

\vspace{5mm}

Even though we have a propensity for the self-shadowing scenario, we cannot firmly establish the origin of the polarized-light deficit in the disk inner region. We argue that in the case of disk scale height enhancement or dust properties change, deeper polarized-light images may still be able to detect emission and its distribution will be unchanged. Conversely, in the dust deficit scenario no significant emission from the inner regions is expected anyway. Then, variable self-shadowing might be confirmed by revealing that the location of the inner edge of the emission is time dependent. Finally, higher resolution images of the DCO+ emission are necessary to constrain the first location of the CO freeze-out.

\section{Discussion: non-detections} \label{Discussion_nondetection}
The disks around HD141569A and HD150193A both show prominent and extended structures out to hundreds of AU in scattered light (see below for references). Therefore, their non-detection in polarized light in the inner $\sim100$ AU provides important information about their structure. In this section, we discuss whether these non-detections are still consistent with previous observations, and speculate on their disk geometry.

\subsection{HD141569A}
\begin{figure}
   \centering
 \includegraphics[width=9cm]{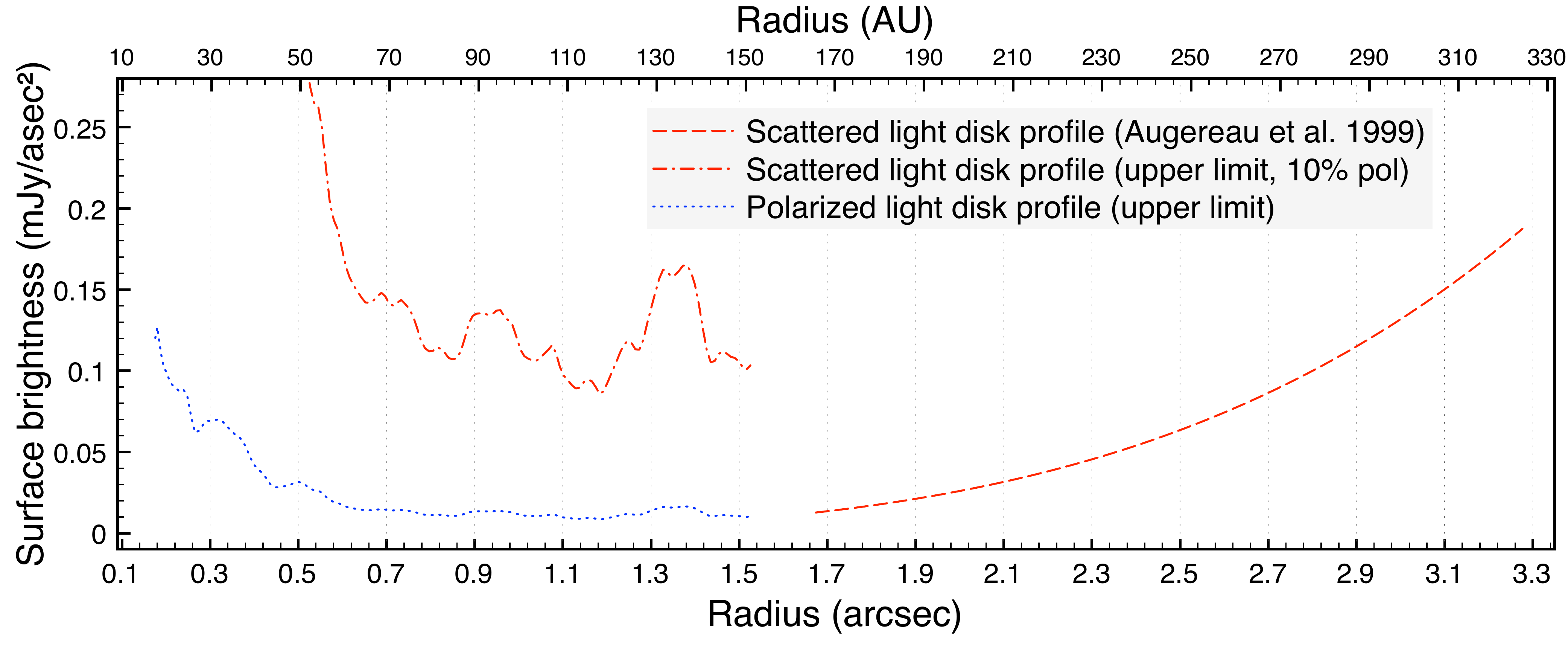} 
 \includegraphics[width=9cm]{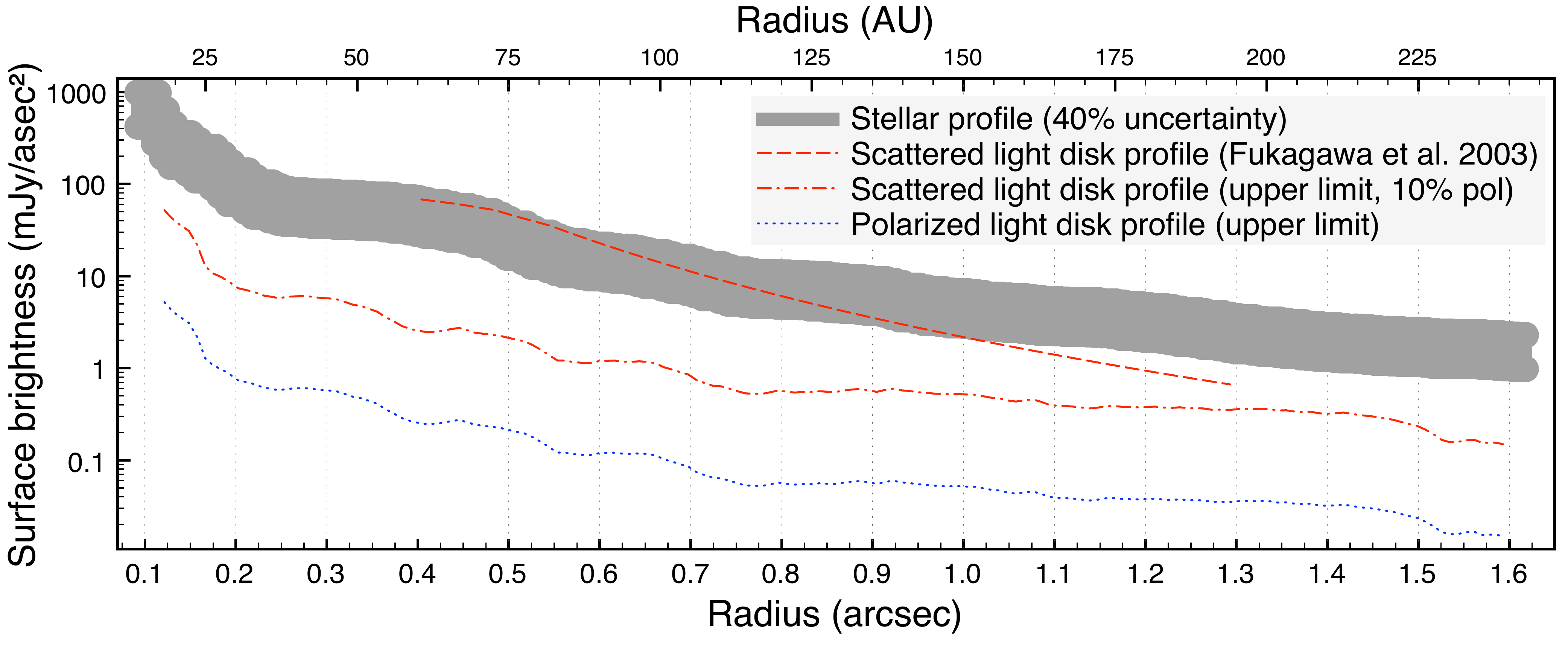}   
  \caption{Surface brightness profile of HD141569A in the $H$ band along the southern major axis (top panel) and of HD150193A in the $H$ band along the northern major axis (bottom panel). The $3\sigma$ upper limit on the polarized-light profile is obtained from the $Q_{\rm T}$ and $U_{\rm T}$ images. The $3\sigma$ upper limit on the scattered-light profile is obtained by assuming a conservative value for the polarization fraction of $10\%$. The gray area in the bottom panel denotes our measured stellar intensity profile, taking into account the $40\%$ uncertainty from the photometric calibration. The scattered-light disk profile by \citet{Fukagawa2003} is still consistent with being an unsubtracted stellar PSF.}
            \label{Radial_nondetections}
  \end{figure}

A disk inner edge of a few hundreds of AU was inferred around HD141569A by \citet{Sylvester1996} by means of SED fitting. Even though this technique is known to be highly degenerate, the deficit of near- to -mid-IR flux shown by the source is effectively suggestive of a large dust gap. Furthermore, $HST$ scattered-light images at 1.1 $\mu$m \citep{Weinberger1999} and 1.6 $\mu$m \citep{Augereau1999} revealed a dramatic decrease of flux inside 160 and 250 AU, respectively, probably indicating a region depleted of material. In the top panel of Fig.\,\ref{Radial_nondetections} we plot the best fit for the scattered-light profile in $H$ band along the southern major axis from \citet{Augereau1999} and compare it with the $3\sigma$ sensitivity of our PDI images. If the scattered-light profile observed with $HST$ does not dramatically increase inward of $1.5\arcsec$, our PDI non-detection is still consistent with it (regardless of the polarization fraction). In the plot we also show the upper limit on the scattered-light profile inside $1.5\arcsec$ assuming a conservative polarization fraction of $10\%$. 

Alternatively, we inspect whether the optical depth inferred at $r<1\arcsec$ by \citet{Marsh2002} would be sufficient to provide a detectable polarized flux. These authors estimated the optical depth $\tau\, (1.1\, \mu {\rm m})$ normal to the disk mid plane, from their mid-IR observations. Interestingly, these values ideally reconnect to those obtained from scattered light farther out \citep{Weinberger1999}. Assuming that $\tau\, (1.1\, \mu {\rm m})\simeq \tau\, (1.6\, \mu {\rm m})$, we consider that for optically thin dust, the optical depth is related to the observed polarized flux through
\begin{equation}
\tau \cdot (A_{\rm scat} \cdot f_{\rm pol})=F_{\rm pol}\cdot 4\pi r^2/F_*
,\end{equation}
where $A_{\rm scat}$ is a quantity that depends on both the intrinsic dust albedo and the scattering function $f_{\rm scat}(\theta)$. By using our upper limit on $F_{\rm pol}$ and the 2MASS $H$ band magnitude \citep{Cutri2003} for $F_*$, we obtain that the condition for our non-detection is $A_{\rm scat} \cdot f_{\rm pol}\lesssim 0.025$. If we assume a conservative $f_{\rm pol}= 10\%$, $A_{\rm scat}$ must be lower than $\sim 0.25$, whereas this parameter can easily be as low as $\sim0.1$ \citep[e.g.,][]{Mulders2013}. 

The presence of dust in the inner 10$-$30 AU was also inferred by \citet{Moerchen2010} through mid-IR imaging. However, forsterite observations suggest that the dust grains in this region are large {\citep[$>10\ \mu$m,][]{Maaskant2014b}}, and thus elusive in near-IR scattered light. Finally, \citet{Maaskant2014} showed that the ionization fraction of polycyclic aromatic hydrocarbons (PAHs) from this source is extremely high, suggesting that they are located in an optically thin environment. 

Therefore, our non-detection of the disk at radii $0.1\arcsec-1\arcsec$ is consistent with earlier observations, and in particular with scattered light from the outer disk \citep{Augereau1999, Weinberger1999} and with mid-IR observations in the inner regions \citep{Marsh2002}. Deeper observations may reveal a polarized signal from the inner disk. 
   
\subsection{HD150193A}
Evidence for the existence of a disk around HD150193A is provided, among others, by the infrared excess \citep{Sylvester2000} and by amorphous and crystalline silicate features \citep{Meeus2001}. 

\citet{Fukagawa2003} revealed a bright $\sim$190-AU-large disk in $H$ band scattered light using Subaru/CIAO. Since the coronagraphic mask was smaller than our OWA, we compared their dataset with our results over a large range of radii ($0.4\arcsec$ to $1.3\arcsec$, see bottom panel of Fig.\,\ref{Radial_nondetections}). To do that, we considered our 3$\sigma$ sensitivity and used it as upper limit. The scattered-to-polarized brightness ratio varies with increasing distance from $\gtrsim270$ to $\gtrsim20$. This yields an upper limit for the polarization fraction of $0.4\%$ at $r=0.5\arcsec$ and $2.4\%$ at $r=1\arcsec$, which is much lower than any value observed or predicted so far for disks. We also set an upper limit on the scattered-light profile by assuming a conservative polarization fraction of $10\%$. Moreover, in Fig.\,\ref{Radial_nondetections} we show that the stellar intensity profile from our dataset is only marginally higher than the disk scattered-light images from \citet{Fukagawa2003}. We estimate that our calibration is accurate to $40\%$ and that any discrepancy by more than a factor of 2 is hard to explain. \citet{Fukagawa2010} noted that their observations may have been affected by the limited quality of the PSF and that further confirmation would be important. Hence, we are more inclined to believe that our upper limits are correct.

In addition, no sign of extended polarized emission was resolved by \citet{Hales2006} in the J band, even though the authors revealed a high degree of polarization, probably due to polarizing material in the line of sight. \citet{Dominik2003} argued that the SED of this source may be explained with a very small disk (with an outer radius as small as 8 AU). Such a compact disk would not be visible to our PDI images, which have an IWA of 15 AU.

Given the above points, the existence of an extended (hundreds of AU) disk is still to be confirmed. A variable self-shadowing scenario may in principle reconcile our dataset and that by \citet{Fukagawa2003}. As discussed in Sect.\,\ref{Discussion_self-shadowing}, a variable inner wall scale height can provide a variable near-IR flux and a time-dependent shadowing effect on the outer disk. This mechanism is more likely to occur in group II objects. Since HD150193A is a group II object, this phenomenon may be the cause of our non-detection and may also explain the previous detection by \citet{Fukagawa2003}. 

If, at the moment of our observations, the disk inner wall was in a "high" state (to cast the speculated shadow), then our total intensity images might be in a high state as well. In fact, a significant fraction of the $H$ -band emission from Herbig Ae/Be stars is thought to originate from the wall itself. Therefore, one may expect high $H$ intensity when the shadow is cast, and vice versa. However, from the bottom panel of Fig.\,\ref{Radial_nondetections} it is clear that our intensity profile will be lower than that from \citet{Fukagawa2003} because it is comparable with their scattered-light profile after PSF subtraction. This finding weakens the possible agreement of the two datasets.

\subsection{Polarized light vs flaring angle}
If the non-detection of HD150193A is motivated by the flat nature of this disk (in contrast to many other bright disks), one may expect a correlation between the amount of (polarized) scattered light and the flaring angle. The disk flaring angle can be estimated through mid-IR photometry, and in particular through the flux ratio at wavelengths of 30 $\mu$m and 13.5 $\mu$m {\citep{Maaskant2013, Maaskant2014b}}. In this approach, the transition from group I to group II objects can be set to $F_{30}/F_{13.5}=2.1$ \citep{Khalafinejad2014}.

\begin{figure}
   \centering
 \includegraphics[width=9cm]{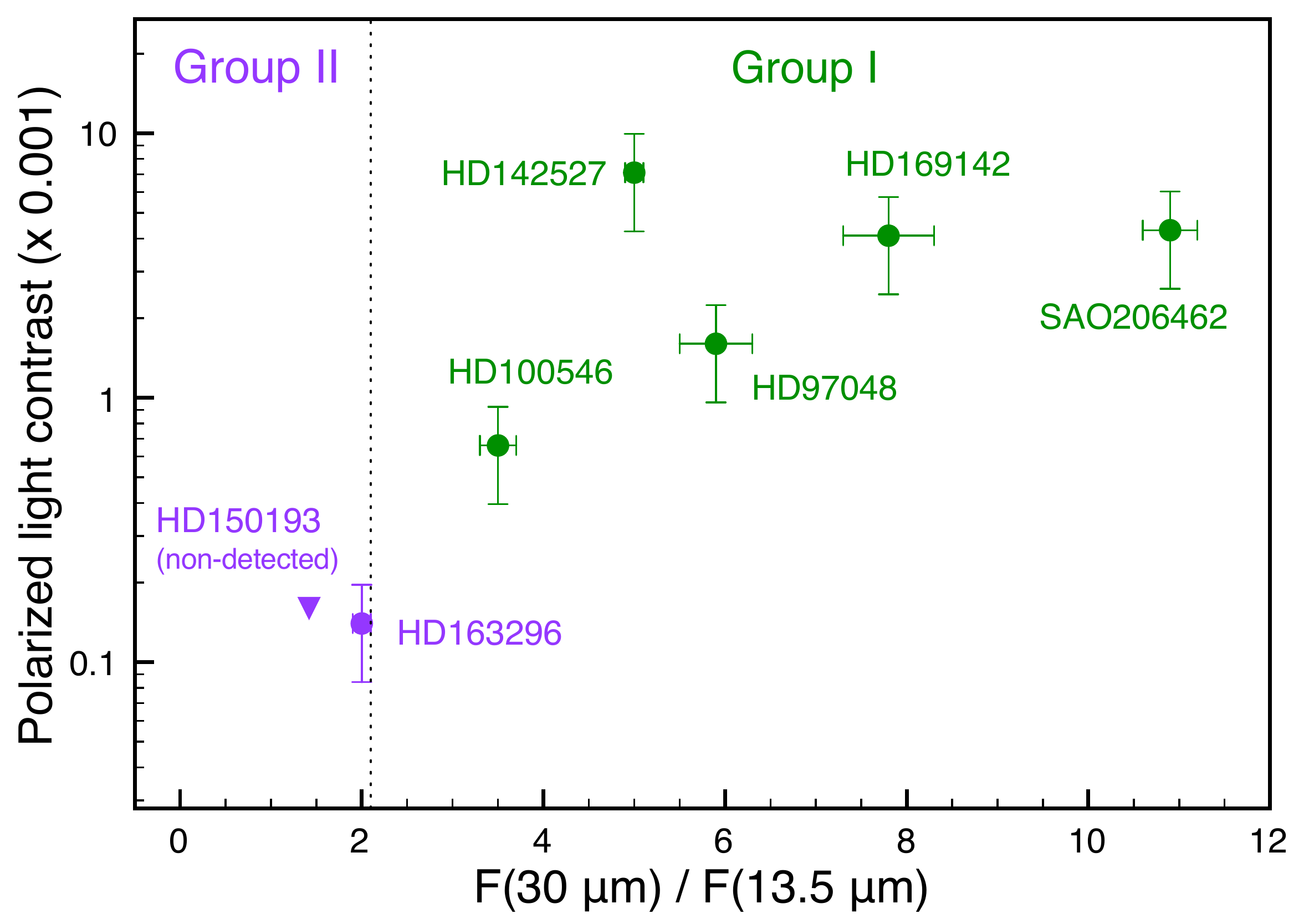}  
  \caption{Polarized-to-stellar light contrast for a sample of Herbig Ae/Be stars compared with the flux ratio at wavelengths of 30 $\mu$m and 13.5 $\mu$m. This ratio is indicative of the disk flaring angle. All group I objects have been clearly detected in polarized light, whereas HD163296 (group I-II) is only weakly detected and HD150193A (group II) is not detected at all. Flux ratios are from \citet{Acke2010}, contrasts from SAO206462, \citet{Garufi2013}; HD169142, \citet{Quanz2013}; HD97048, \citet{Quanz2012}; HD142527, \citet{Avenhaus2014}; HD100546, \citet{Quanz2011}; HD163296, and HD150193A (this work).}
            \label{Contrast_color}
  \end{figure}

In Fig.\,\ref{Contrast_color} we show that a correlation between the polarized-light contrast and the flaring angle holds for all objects from this or from our previous campaigns. To do that, we used the mid-IR photometry from \citet{Acke2010} and the polarized-to-stellar light brightness contrast from Eq.\,\ref{Formula_albedo}. From the plot, it is clear that all objects that were clearly resolved in PDI are flared (group I objects). Conversely, HD163296 (marginally detected in PDI) lies at the edge of the two groups and HD150193A (non-detected) is even below this formal threshold. This therefore supports the idea that the non-detection of HD150193A and the weak detection of HD163296 are due to their particular disk geometry
and not to any peculiar dust properties therein.

\section{Summary} \label{Conclusions}
We have presented new $H$ and $K_{\rm S}$ PDI observations of HD163296, HD141569A, and HD150193A. These objects were known to host very extended disk structures. Our observations only revealed a faint or a totally absent contribution in polarized light. In particular, the disk around HD163296 is weakly detected in both bands. The brightness distribution is strongest along the major axis, and only a tentative detection is present along the minor axis. The radial profile smoothly decreases to the noise level inward of $\sim0.6\arcsec$. The disk emission is likely to be redder than the stellar emission. On the other hand, HD141569A and HD150193A were not detected in any band.

The incongruity between these and previous observations inspire a discussion on the geometry of each object.

\begin{itemize}
\item HD163296: the peculiar azimuthal distribution of the polarized light from our images does not allow us to solidly constrain the morphology of dust grains. Asymmetries and colors may suggest that the disk surface is dominated by sub-$\mu$m size dust grains, but deeper photometrically reliable observations are needed to confirm this scenario. The radial profile is somewhat surprising, since the object is not known to be a transition disk. We argue that the favored explanation for the inner light drop is a (variable) self-shadowing. Since the disk is a border-line object between group I and II, a scenario where the disk comes out of the shadow-cone at $\sim70$ AU only is realistic. Near-IR photometry and scattered-light variability from previous works support this hypothesis. Three other scenarios, however, may potentially generate the same light drop: a dramatic change in the disk scale height, a notable dust grain growth, and an enhanced dust albedo due to CO molecule freeze-out. 

\item HD141569A: given our sensitivity, our non-detection of the disk from 10 to 100 AU is still consistent with scattered-light images of the external disk, which show a significant flux deficit inward of $\sim200$ AU. The dust optical depth inferred from mid-IR imaging is also consistent with our non-detection. Slightly deeper observations may still detect a contribution from the inner 100 AU.

\item HD150193A: this non-detection is not consistent with earlier scattered-light images, which might be affected by inaccurate PSF subtraction. The disk may be smaller than our inner working angle or self-shadowed. The latter scenario, in particular, is supported by the flat nature of the disk. Deeper observations might not resolve any contribution from the disk.

\end{itemize}

We have also shown a significant correlation between the amount of polarized scattered light and the flaring angle for a number of Herbig Ae/Be disks from our PDI campaigns. This highlights the importance of knowing the disk geometry (as suggested by, e.g., SEDs) in the interpretation of scattered-light images. It also suggests a potential threshold in the disk flaring angle for the detection of these objects in scattered light, as well as the need for deeper observations when imaging flatter objects. A possible analogous correlation for T Tauri disks would allow us to draw conclusions on the (dis)similarities of these two families of disks.

Finally, the time-dependence of the inner disk geometry motivates the importance of multi-epoch observations in scattered light. In particular, these could unravel the process of self-shadowing in Herbig Ae/Be disks. In this context, it is particularly important to monitor any potential correlation between the total near-IR intensity and the simultaneous scattered-light brightness. 

\begin{acknowledgements}
The authors acknowledge the staff at VLT for their excellent support during the observations. We thank the referee for valuable comments, Geoff Mathews for sharing the ALMA data on HD163296, and Koen Maaskant for stimulating discussions. This research has made use of the SIMBAD database, operated at CDS, Strasbourg, France.
\end{acknowledgements}

\bibliographystyle{aa} 
\bibliography{Reference.bib} 

\end{document}